\documentclass[10pt,conference]{IEEEtran}
\IEEEoverridecommandlockouts

\newif\ifDEBUG
\DEBUGfalse

\newcommand{\totalPapersFound}{17,718\xspace}               
\newcommand{\papersWithOpenAlex}{17,364\xspace}             
\newcommand{\papersWithCitations}{14,998\xspace}            
\newcommand{\candidateNaturalSciencePapers}{7,212\xspace}   
\newcommand{\markdownPapers}{6,956\xspace}                  

\newcommand{\bioProp}{36.91\%\xspace}
\newcommand{\envProp}{19.31\%\xspace}
\newcommand{\neuroProp}{18.59\%\xspace}
\newcommand{\agProp}{12.18\%\xspace}
\newcommand{\earthProp}{4.09\%\xspace}
\newcommand{\immProp}{3.48\%\xspace}
\newcommand{\physProp}{3.19\%\xspace}
\newcommand{\chemProp}{2.26\%\xspace}

\newcommand{\papersUsingDeepLearning}{4,656\xspace}                                         
\newcommand{\papersUsingDeepLearningProp}{66.94\%\xspace}                                   
\newcommand{\openalexDLFieldOne}{``Biochemistry, Genetics and Molecular Biology''\xspace}   
\newcommand{\openalexDLFieldOneCount}{2,015\xspace}                                         
\newcommand{\openalexDLFieldTwo}{``Neuroscience''\xspace}                                   
\newcommand{\openalexDLFieldTwoCount}{1,199\xspace}                                         
\newcommand{\openalexDLFieldThree}{``Environmental Science''\xspace}                        
\newcommand{\openalexDLFieldThreeCount}{899\xspace}                                       

\newcommand{\papersResuingPTMs}{1,833\xspace}                                               
\newcommand{\papersResuingPTMsProp}{39.36\%\xspace}                                         
\newcommand{\openalexPTMFieldOne}{``Biochemistry, Genetics and Molecular Biology''\xspace}  
\newcommand{\openalexPTMFieldOneCount}{888\xspace}                                        
\newcommand{\openalexPTMFieldTwo}{``Neuroscience''\xspace}                                  
\newcommand{\openalexPTMFieldTwoCount}{416\xspace}                                          
\newcommand{\openalexPTMFieldThree}{``Agricultural and Biological Sciences''\xspace}        
\newcommand{\openalexPTMFieldThreeCount}{357\xspace}                                        

\newcommand{\topImpactOne}{``testing''\xspace}

\newcommand{\onePTMReusePattern}{``Adaptation''\xspace}         
\newcommand{\onePTMReusePatternLower}{``adaptation''\xspace}    
\newcommand{\ptmAdaptationReuse}{1,329\xspace}                  
\newcommand{\ptmAdaptationReuseProp}{70.29\%\xspace}            
\newcommand{\ptmDeploymentReuse}{318\xspace}                    
\newcommand{\ptmDeploymentReuseProp}{17.31\%\xspace}            
\newcommand{\ptmConceptualReuse}{50\xspace}                     
\newcommand{\ptmConceptualReuseProp}{2.72\%\xspace}             

\newcommand{\impactAnalysis}{265\xspace}
\newcommand{\impactAnalysisProp}{14.42\%\xspace}
\newcommand{\impactBackground}{115\xspace}
\newcommand{\impactBackgroundProp}{6.26\%\xspace}
\newcommand{\impactObservation}{5\xspace}
\newcommand{\impactObservationProp}{0.27\%\xspace}
\newcommand{\impactTest}{1,123\xspace}
\newcommand{\impactTestProp}{61.13\%\xspace}
\newcommand{\impactHypothesis}{55\xspace}
\newcommand{\impactHypothesisProp}{2.99\%\xspace}

\newcommand{\analysisDate}{March 18th, 2026\xspace}
\newcommand{\analysisEndDate}{December 31st, 2025\xspace}
\newcommand{\analysisStartDate}{January 1st, 2000\xspace}
\usepackage{balance}
\usepackage{orcidlink}
\usepackage{xurl}

\usepackage[most]{tcolorbox}  

\usepackage{listings}
\usepackage{xcolor}  

\usepackage{afterpage}
\usepackage{algorithmic}
\usepackage{array}
\usepackage{booktabs}
\usepackage[skip=4pt]{caption}
\usepackage[compress]{cite}
\usepackage{enumitem}
\usepackage{float}
\usepackage{fontawesome5}
\usepackage[T1]{fontenc}
\usepackage{graphicx}
\usepackage{lipsum}
\usepackage{lscape}
\usepackage{makecell}
\usepackage{multirow}
\usepackage{siunitx}
\usepackage{soul}
\usepackage{rotating}
\usepackage{tabularx}
\usepackage{longtable}
\usepackage{tcolorbox}
\usepackage{textcomp}
\usepackage{titlesec}
\usepackage[normalem]{ulem}
\usepackage{xspace}
\usepackage{balance}

\usepackage{hyperref}
\usepackage[nameinlink,noabbrev]{cleveref}
\usepackage[numbers,sort&compress]{natbib}

\newcommand{\ie}{\textit{i.e.},\xspace}  
\newcommand{\eg}{\textit{e.g},\xspace}   

\newcommand{\myparagraph}[1]{\vspace{0.20cm}\noindent\textbf{#1} \noindent{}}   

\newcounter{finding}                

\ifDEBUG
    \newcommand{\TODO}[1]{\hl{TODO: #1}}    
    \newcommand{\CITE}[1]{\hl{CITE: #1}}    
    
    \newcommand{\NMS}[1]{\textcolor{orange}{[NMS: #1]}} 
    \newcommand{\GKT}[1]{\textcolor{green}{[GKT: #1]}}  
    \newcommand{\JD}[1]{\textcolor{blue}{[JD: #1]}}     
    \newcommand{\KL}[1]{\textcolor{teal}{[KL: #1]}}     
\else
    \newcommand{\TODO}[1]{}
    \newcommand{\CITE}[1]{}

    \newcommand{\NMS}[1]{}
    \newcommand{\GKT}[1]{}
    \newcommand{\JD}[1]{}
    \newcommand{\KL}[1]{}
\fi

\begin{document}

\title{An Empirical Investigation of Pre-Trained Deep Learning Model Reuse in the Scientific Process}



\author {
    \IEEEauthorblockN   {
        Nicholas M. Synovic\,\orcidlink{0000-0003-0413-4594}
        \IEEEauthorrefmark{1}%
        \thanks {
            Corresponding authors: Nicholas M. Synovic
            (\texttt{nsynovic@luc.edu}) and George K. Thiruvathukal
            (\texttt{gthiruvathukal@luc.edu}).
        },%
        \quad
        Karolina Ryzka\,\orcidlink{0009-0000-2009-5957}
        \IEEEauthorrefmark{1},%
        \quad
        Alessandra V. Vellucci Solari\,\orcidlink{0009-0007-2686-1668}
        \IEEEauthorrefmark{1},\\%
        \quad
        Kenny Lyons\,\orcidlink{0009-0009-1873-7960}
        \IEEEauthorrefmark{1},%
        \quad
        James C. Davis\,\orcidlink{0000-0003-2495-686X}
        \IEEEauthorrefmark{2},%
        \quad
        and~George K. Thiruvathukal\,\orcidlink{0000-0002-0452-5571}
        \IEEEauthorrefmark{1}%
  }

    \IEEEauthorblockA   {
        \IEEEauthorrefmark{1}%
        \textit{Computer Science Department, Loyola University Chicago}\\
        Chicago, IL, USA
    }

    \IEEEauthorblockA   {
        \IEEEauthorrefmark{2}%
        \textit{Electrical and Computer Engineering Department, Purdue University}\\
        West Lafayette, IN, USA
    }
}

\maketitle

\def\BibTeX{{\rm B\kern-.05em{\sc i\kern-.025em b}\kern-.08em
    T\kern-.1667em\lower.7ex\hbox{E}\kern-.125emX}}

\begin{abstract}
Deep learning has achieved recognition for its impact within natural sciences, yet the prohibitive financial and technical cost of training models from scratch inhibits adoption.
Following software engineering community guidance, natural scientists are reusing pre-trained deep learning models (PTMs) to amortize these costs.
While prior works recommend PTM reuse patterns, we present the first empirical study of these patterns in the natural sciences, quantifying the utilization and impact of PTM reuse within the scientific process across ~\totalPapersFound peer reviewed, open access papers.
Our results show that
    \openalexPTMFieldOne has outpaced other natural scientific fields in PTM reuse,
    \onePTMReusePatternLower reuse is the most prevalent PTM reuse pattern identified across all natural science fields, and
    the~\topImpactOne stage of the scientific process has been most impacted by PTM integration.
\end{abstract}


\begin{figure}[h]
    \centering
    \includegraphics[width=\linewidth]{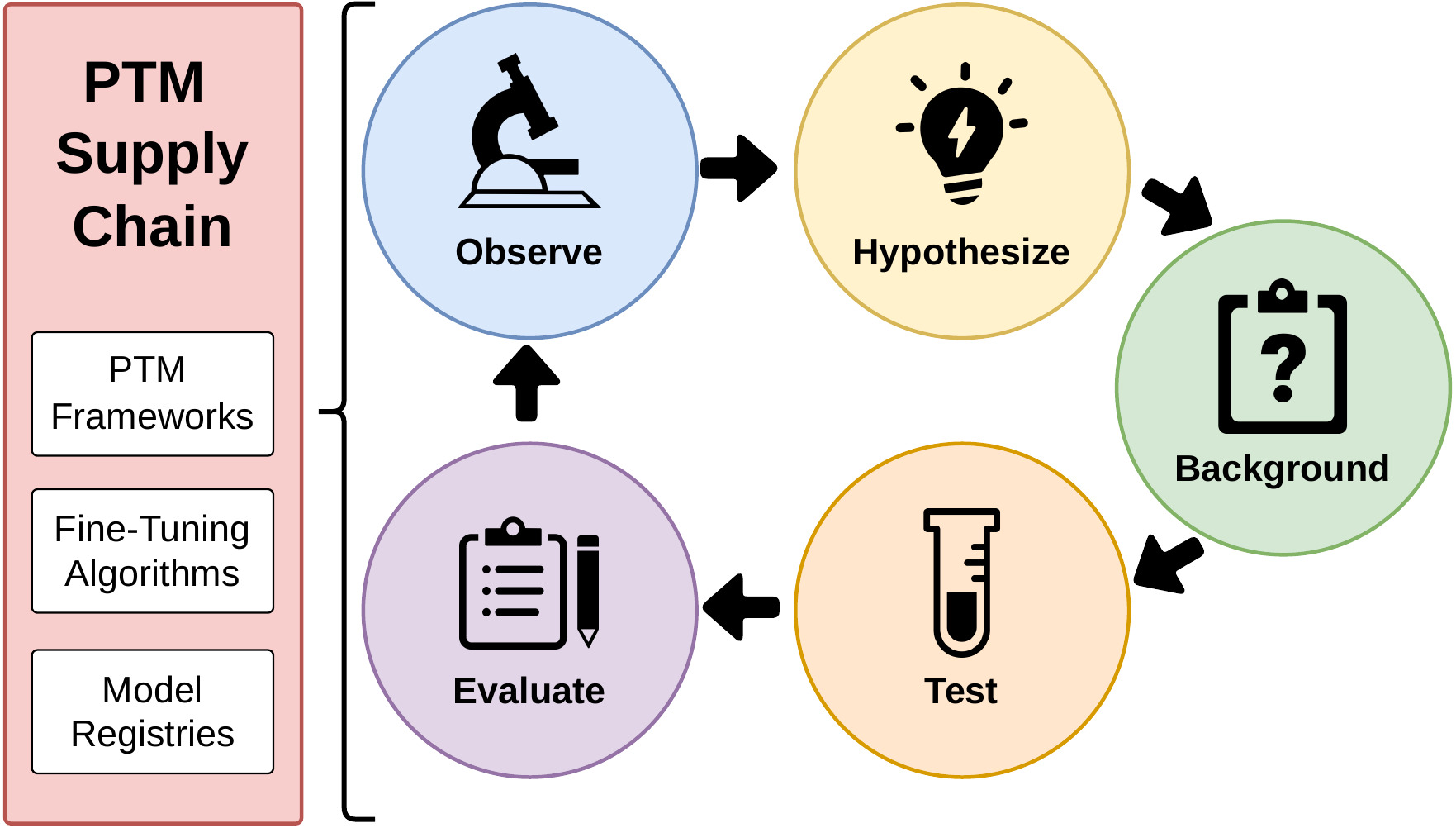}
    \caption{
    PTMs support every stage of the scientific process --- from observation and hypothesis generation through testing and evaluation. 
    A supply chain of 
        frameworks, 
        fine tuning algorithms, and 
        model registries 
    enables scientists to apply data driven computational methods in their research.
    }
    \vspace{-15pt}
    \label{fig:fig1}
\end{figure}

\section{Introduction}
The 2024 Nobel Prizes in 
    Physics~\citep{the_royal_swedish_academy_of_sciences_nobel_physics_2024} and 
    Chemistry~\citep{the_royal_swedish_academy_of_sciences_nobel_chemistry_2024} 
honor 
    foundational neural networks~\citep{hopfield_neural_1982} and 
    their transformative application to protein structure prediction~\citep{jumper_highly_2021}, 
validating the role of deep learning (DL) in science.
DL models enable scientists to extract insights from natural systems~\citep{lecun_deep_2015}, frequently matching or exceeding the performance of hand crafted methods~\citep{abramson_accurate_2024}.

While DL offers transformative opportunity for the natural sciences, prohibitive costs inhibit creating project specific models. 
Pre-trained DL models (PTMs) amortize these costs~\citep{han_pre-trained_2021, davis_reusing_2023, jiang_empirical_2023}, but reusing them is non-trivial due to 
    algorithmic complexity~\citep{tang_empirical_2021}, 
    dependence on data quality~\citep{sculley_hidden_2015}, and 
    challenges in deploying and maintaining models~\citep{menshawy_navigating_2024}. 
Prior works have proposed qualitative strategies to address these challenges~\citep{carver_software_2016, lecun_deep_2015, ching_opportunities_2018, mater_deep_2019, tanaka_deep_2021}, however, to our knowledge, no prior work has empirically quantified the prominence and impact of PTM reuse patterns~\citep{davis_reusing_2023, jiang_empirical_2023} within the natural sciences.

To address this, we conduct the first empirical investigation of PTM reuse within the natural sciences.
Our contributions are: 
    (1) systematic identification of PTM reuse patterns via an automated, large language model (LLM) driven pipeline, 
    (2) empirical characterization of how these models augment the scientific workflow, and 
    (3) a dataset mapping scientific literature to specific PTM reuse patterns and their workflow impacts.

By analyzing~\totalPapersFound peer reviewed, open access papers from four megajournals~\citep{bjork_evolution_2018}, we identified~\papersResuingPTMs natural science papers reusing PTMs. 
For each paper, we then mapped it to 
    their reused PTMs, 
    reuse patterns per PTM, and 
    the impacted scientific process step \NMS{workflow stage -> process step}. 
\openalexPTMFieldOne emerged as the dominant field for PTM adoption, 
    the~\onePTMReusePatternLower pattern was the predominant mode of PTM reuse, and 
    the~\topImpactOne stage experienced the most significant impact.

LLM prompts, dataset, and source code to replicate this study are availible via this Zenodo artifact~\citep{synovicEmpiricalInvestigationPreTrained2026aZenodo}.

\section{Background and Related Work}\label{sec:background}
Pre-trained deep learning model (PTM) reuse in the natural sciences draws on established software engineering reuse patterns and their integration into the scientific workflow.

\subsection{Pre-Trained Model Reuse}\label{sec:background-ptm_reuse} 
\begin{table}[t]
    \caption{
    PTM reuse patterns identified in our study, with definitions and representative papers leveraging each pattern.
    }
    \begin{tabular}{p{0.19\linewidth} p{0.47\linewidth} p{0.17\linewidth}}
        \hline
        \textbf{Reuse Pattern} & \multicolumn{1}{c}{\textbf{Definition}} & \textbf{Examples} \\ 
        \hline
        Conceptual & Replicate or reengineer PTMs described in academic literature. & \citep{lachance_practical_2020, pan_generalizing_2023, dutagaci_characterization_2023} \\
        Adaptation & Modify existing PTMs to address different tasks. & \citep{rattray_machine_2023, ha_classification_2023, cai_detection_2021} \\
        Deployment & Engineer existing PTMs for use in different computational frameworks. & \citep{dick_differential_2020, oyarbide_sperm_2023, mathew_genome_2019} \\ 
        \hline
    \end{tabular}
    \label{table:tableH}
    \vspace{-10pt}
\end{table}
Deep learning (DL) models are emerging as an increasingly important dependency in software systems~\citep{tang_empirical_2021, jiang_peatmoss_2024, jiang_empirical_2022}. 
However, training these models incurs significant 
    technical~\citep{sculley_hidden_2015}, 
    financial~\citep{schwartz_green_2020}, and 
    environmental~\citep{strubell_energy_2020, schwartz_green_2020} 
costs. 
To mitigate these, engineers leverage PTMs, which encapsulate a model's architecture and learned weights into a distributable package~\citep{davis_reusing_2023}. 
This is facilitated by an ecosystem of DL 
    frameworks~\citep{wolf_huggingfaces_2020, ansel_pytorch_2024}, 
    fine tuning algorithms~\citep{prottasha_peft_2025}, and 
    PTM registries~\citep{huggingface_2024, yu_tensorflow_2020} 
that provide hosting infrastructure and programmatic interfaces for model management (\Cref{fig:fig1}). 
Within this ecosystem, PTM reuse implementations can be categorized into the distinct patterns~\citep{davis_reusing_2023} listed in~\Cref{table:tableH}.

While prior work has interviewed practitioners to understand PTM applications and reuse challenges~\citep{jiang_challenges_2024, jiang_empirical_2023}, empirical efforts remain focused on software systems, including 
    investigations into conceptual reuse portability~\citep{jiang_challenges_2024}, 
    taxonomies of adaptation techniques~\citep{prottasha_peft_2025}, and 
    deployment difficulties~\citep{jajal_analysis_2023}. 
To our knowledge, this work is the first empirical study of self reported PTM reuse practices among researchers applying PTMs in the natural sciences. 

\subsection{Pre-Trained Models in Science}\label{sec:background-ptm4sci}
DL methods are increasingly leveraged in the natural sciences to enable data driven discoveries~\citep{lecun_deep_2015, duede_oil_2024, angermueller_deep_2016, ramsundar_deep_2019, mater_deep_2019, tanaka_deep_2021, bergen_machine_2019}.
These methods have assisted in 
    molecular protein design~\citep{zambaldi_novo_2024, krishna_generalized_2024, jumper_highly_2021}, 
    understanding chemical reactions~\citep{schwaller_extraction_2021, zhang_large_2025}, and 
    modeling seismic events~\citep{wang_deep_2025, mousavi_deep-learning_2022}.
Current recommendations favor refinement of existing PTMs on domain specific data over costly development of bespoke models~\citep{lee_ten_2022}. 
However, to our knowledge, little work has evaluated how PTMs are being integrated into the scientific process.

Our work contributes to formalizing PTM reuse within the natural sciences by empirically evaluating established reuse patterns and their impact on the scientific process. 
By mapping how PTMs are integrated, we provide a framework for assessing their impact on discovery and informing more robust and reproducible software ecosystems for scientific inquiry.
Our work extends existing software engineering foundations by characterizing how scientists employ 
    ``adaptation'', 
    ``conceptual'', and 
    ``deployment'' 
PTM reuse patterns within the scientific process, identifying both prevalent practices and underused opportunities.

\section{Research Questions}\label{sec:rqs}
\begin{figure}[t]
    \centering
    \includegraphics[width=\linewidth]{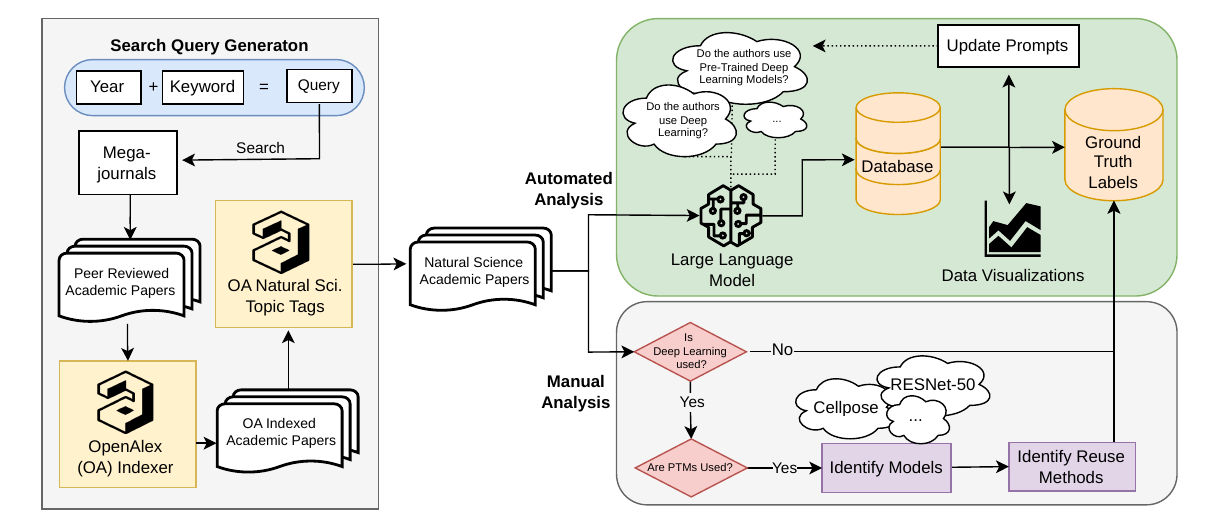}
    \caption{
    Our methodology for identifying and analyzing peer reviewed, open access natural-science publications in megajournals. 
    Publications retrieved from megajournal databases are enriched with OpenAlex metadata and analyzed for 
        DL use, 
        PTM reuse, and 
        PTM reuse patterns. 
    Manually curated ground truth labels guide refinement of the prompting strategy.
    }
    \label{fig:figB}
    \vspace{-15pt}
\end{figure}
Scientists studying 
    biology, 
    chemistry, 
    physics, and 
    environmental science 
increasingly reuse pre-trained deep learning models (PTMs) to support data driven discovery. 
However, little is known about how PTM reuse patterns have evolved over time or how PTMs are used across the steps of the scientific process. 
We therefore ask:

\begin{itemize}[leftmargin=26pt, rightmargin=5pt]
    \item[\textbf{RQ1}] What are the longitudinal changes in PTM reuse patterns?
    \item[\textbf{RQ2}] Across which steps of the scientific process are PTMs applied?
\end{itemize}

\Cref{fig:figB} presents our methodology.
We provide the specific methods and empirical results for RQ1 and RQ2 in~\Cref{sec:rq1} and~\Cref{sec:rq2} respectfully.

\section{RQ1: Longitudinal PTM Reuse}\label{sec:rq1}
This section details our methodology for quantifying the adoption and longitudinal evolution of pre-trained deep learning model (PTM) reuse patterns within natural science between~\analysisStartDate and~\analysisEndDate.

\subsection{Methods}\label{sec:rq1-methods}
To identify PTM reuse patterns in the natural sciences, we developed an automated, large language model (LLM) driven pipeline for analyzing scholarly publications. 
First, we retrieve candidate peer reviewed, open access articles from megajournals and normalize their metadata.
Second, we identify eligible natural science articles and preprocess their prose for LLM based analysis. 
Third, we use an LLM to identify evidence of 
    deep learning (DL) use, 
    PTM reuse, and 
    specific PTM reuse patterns.

\subsubsection{Extraction of Candidate Papers From Mega Journals}\label{sec:rq1-methods-extract}
We characterize PTM reuse patterns in the natural sciences by constructing a dataset of peer reviewed, open access publications. 
Existing methodologies for mapping publications to software rely on tracing citations from known software artifacts~\citep{malviya-thakur_scicat_2023}.
However, because scientific software is frequently under cited~\citep{howison_software_2016}, a citation centric approach risks omitting relevant work. 
To mitigate this, we first query open access megajournals for DL keywords and then normalize the resulting metadata with OpenAlex~\citep{priem_openalex_2022}.

\begin{figure}[h]
    \centering
    \includegraphics[width=\linewidth]{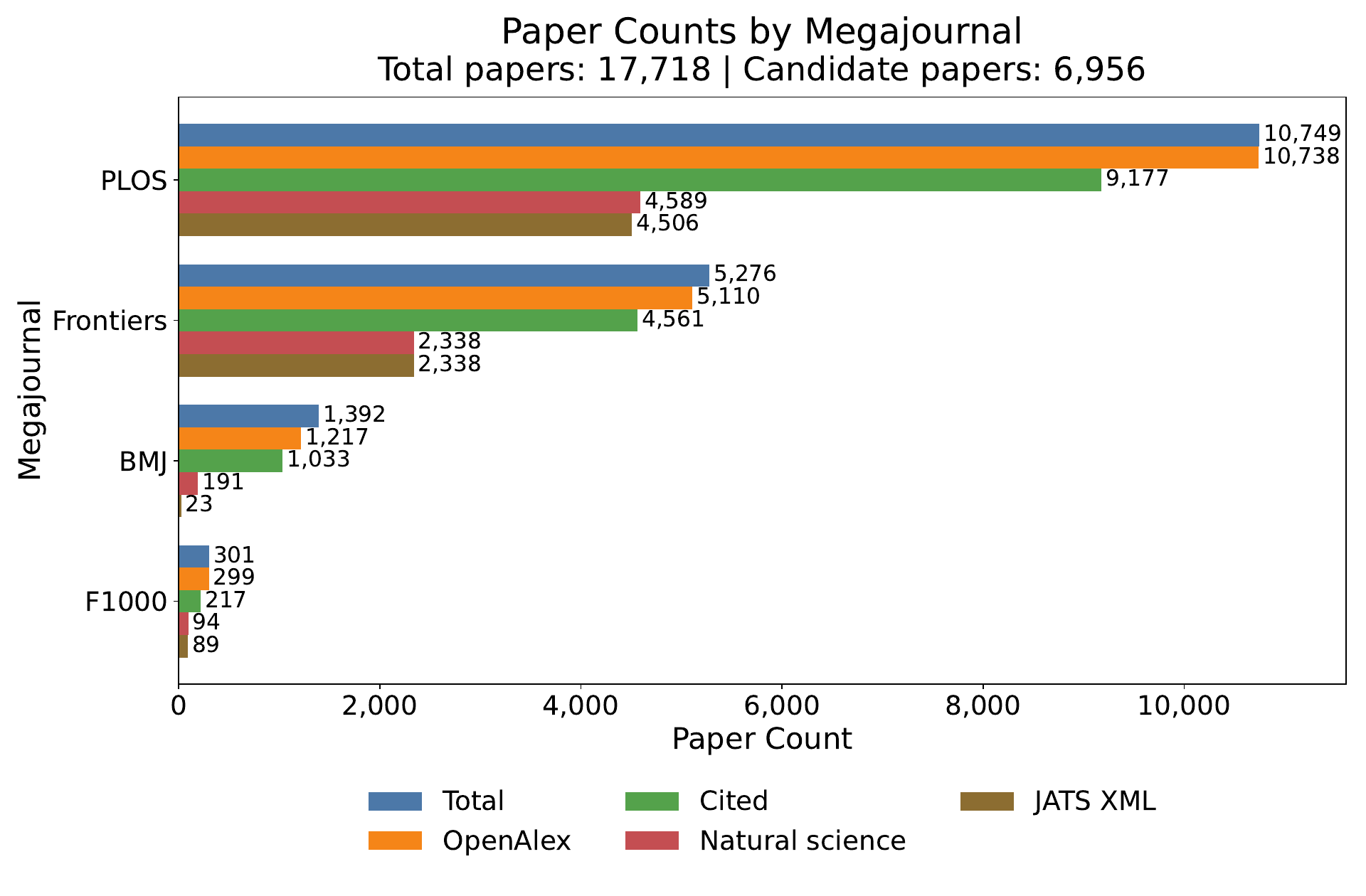}
    \caption{
    Results of the mega-journal paper filtering process. 
    Our queries returned~\totalPapersFound total unique papers between 2000 and 2025, of which~\papersWithOpenAlex where indexed by OpenAlex. 
    Of these,~\papersWithCitations had citations and~\candidateNaturalSciencePapers were natural science papers.
    \markdownPapers of these were available in JATS XML format.
    }
    \label{fig:figQ}
    \vspace{-15pt}
\end{figure}

Our study analyzes peer reviewed, open access papers published between~\analysisStartDate and~\analysisEndDate from the 
    BMJ, 
    F1000, 
    FrontiersIn, and 
    PLOS 
megajournals, serving as representative proxies for broad scientific discourse~\citep{bjork_evolution_2018}. 
We queried each journal using the following keywords: 
    ``Deep Learning'', 
    ``Deep Neural Network'', 
    ``Model Weights'', 
    ``Hugging Face''~\citep{jiang_empirical_2023}, 
    ``Model Checkpoint''~\citep{jiang_empirical_2022}, and 
    ``Pre-Trained Model''~\citep{davis_reusing_2023}. 
The first three are foundational to DL while the remainder were identified from prior work.
This search yielded~\totalPapersFound unique papers as of~\analysisDate. 
\Cref{fig:figQ} presents the total papers returned per megajournal.

We then normalize publication metadata using OpenAlex~\citep{priem_openalex_2022} to ensure a consistent schema for downstream processing on qualitative features (\eg scientific field).
Papers not indexed in OpenAlex were filtered out, resulting in~\papersWithOpenAlex papers for further processing.
Examples of filtered papers are presented in the appendix of our preprint~\citep{synovicEmpiricalInvestigationPreTrained2026}.  

\subsubsection{Filtering and Transformation of Natural Science Papers}\label{sec:rq1-methods-transform}
From the~\papersWithOpenAlex OpenAlex indexed candidate papers, we first filter for papers cited at least once by other works to focus our study on impactful works, yielding~\papersWithCitations papers.
We then filter on OpenAlex topic fields to identify natural science works. 

OpenAlex classifies papers into 26 distinct field topics and returns the top three most likely fields for a given work~\citep{noauthor_oadocumentation_2024}, but does not provide an explicit ``natural science'' classification. 
To address this, we manually applied a binary classification of OpenAlex topic fields as either ``natural science'' or not per~\citep{encyclopaedia_britannica_natural_2025}. 
Two authors independently audited the fields and engaged in inter-author agreement until consensus was reached, resulting in eight natural science topic fields. 
Filtering for works with at least one natural science field from the set of papers cited at least once yielded~\candidateNaturalSciencePapers papers. 
The manually identified fields, their proportions, and example works are presented in~\Cref{table:tableJ}.

We then normalize each paper's prose by retrieving it in JATS XML format~\citep{national_information_standards_organization_jats_standing_committee_ansiniso_2024} if available, removing non-essential content (\eg metadata, linked content, citations) 
and then converting the document to Markdown using \verb|pandoc|~\citep{MacFarlane_Pandoc} for LLM analysis. 
This resulted in~\markdownPapers total papers suitable for evaluation.

\subsubsection{Automated Analysis of Filtered Papers}\label{sec:rq1-methods-load}
To characterize the longitudinal shifts in PTM reuse, we developed an automated extraction pipeline to analyze the candidate natural science works. 
For each publication, we address a four tier hierarchy of inquiries: 
    (1) identification of DL methods, 
    (2) verification of PTM reuse, 
    (3) extraction of specific PTM identifiers, and 
    (4) classification of the associated reuse pattern 
        (\ie 
        adaptation, 
        conceptual,  and 
        deployment). 
Questions (1) and (2) function as exclusionary filters whereby papers failing to meet these criteria are removed from subsequent analysis.

We utilized CO-STAR templated system prompts~\citep{teo_how_2023} to enable OpenAI 
\verb|gpt-5.4-nano-2026-03-17|
to analyze each paper.
To mitigate non-determinism and ensure reproducibility, we set a static seed value of 42.
We set the model to use ``high'' reasoning effort and return ``detailed'' summaries to prioritize complex logical inference.
To enhance the recognition of PTM names, we augmented our prompts with a set of known architectures from the Hugging Face 
    \verb|Transformers|~\citep{wolf-etal-2020-transformers} and 
    \verb|Diffusers|~\citep{von_Platen_Diffusers_State-of-the-art_diffusion}
libraries for questions (2), (3) and (4)
tTo guide system prompt development, 20 LLM responses were validated against manually annotaxted sections.
Prompts were modified until agreement reached 90\% and are detailed in the appendix of our preprint~\citep{synovicEmpiricalInvestigationPreTrained2026} 

\begin{table}[t]        
\caption{The eight identified natural science OpenAlex topic fields, their proportions in our study population, and example papers for each topic.}
        \begin{tabular}{p{0.6\linewidth} p{0.15\linewidth} p{0.1\linewidth}}
        \hline 
            \textbf{OpenAlex Topic Field} & 
            \textbf{Examples} & 
            \textbf{Prop.} \\ 
        \hline 
            Biochemistry, Genetics and Molecular Biology &
            \citep{bio1, bio2, bio3} &
            \bioProp \\
            Environmental Science & 
            \citep{palanichamy_machine_2022, li_fish_2025, guo_integration_2025} & 
            \envProp \\
            Neuroscience & 
            \citep{milyani_deep_2025, umair_privacypreserving_2025, otarbay_svm-enhanced_2025} & 
            \neuroProp \\ 
            Agricultural and Biological Sciences & 
            \citep{ergun_attention-enhanced_2025, salman_plant_2025, tan_p4cn-yolov5s_2025} & 
            \agProp \\ 
            Earth and Planetary Sciences & 
            \citep{jha_deep_2025, kosmala_integrating_2018, miao_using_2018} &
            \earthProp \\ 
            Immunology and Microbiology & 
            \citep{lee_identification_2024, sethna_population_2020, barreto_novel_2024} & 
            \immProp \\ 
            Physics and Astronomy & 
            \citep{peng_multi-scale_2015, mu_ipso-lstm_2023, yang_diffusion_2023} & 
            \physProp \\ 
            Chemistry & 
            \citep{tariq_specollate_2021, gueto-tettay_multienzyme_2023, zhang_multimodal_2024} & 
            \chemProp \\ 
        \hline 
        \end{tabular} 
    \label{table:tableJ}
    \vspace{-10pt}
\end{table}
\subsection{Data Analysis}\label{sec:rq1-analysis}
To identify changes in PTM reuse patterns within the natural sciences, we analyzed trends in publications leveraging DL, PTMs, and PTM reuse patterns over time and by discipline.

\begin{figure*}[htbp]
    \centering
    \begin{minipage}[t]{0.49\linewidth}
        \vspace{0pt}
        \centering      \includegraphics[width=\linewidth]{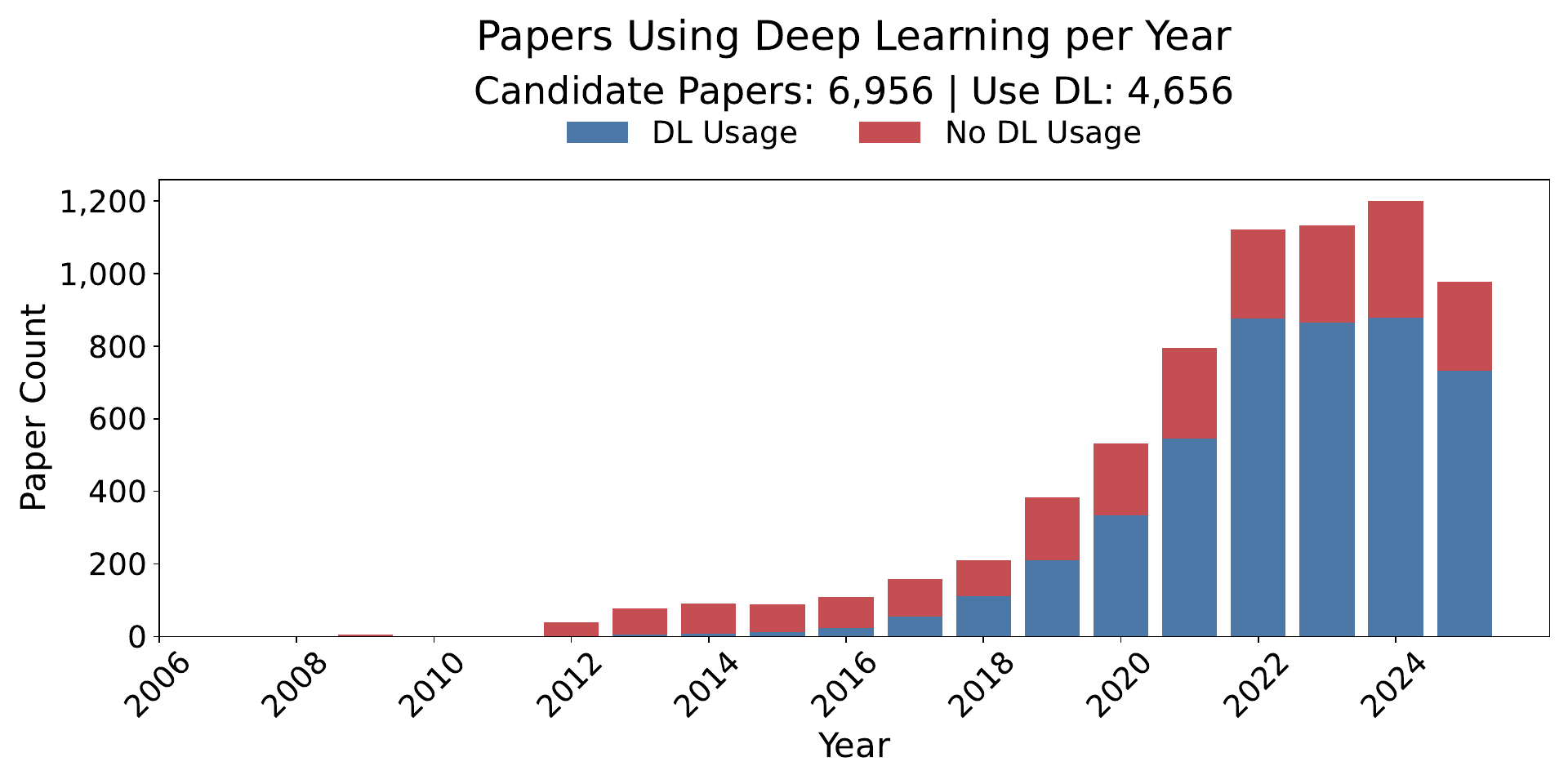}
        \captionof{figure}{
            Number of papers identified using DL methods per year
            plotted against papers that do not use DL.
            The earliest identified paper that uses DL was published
            in 2009~\citep{10.1371/journal.pone.0006393}, however it was not
            until 2012~\citep{qi_unified_2012} that the number of papers
            leveraging DL methods began increasing yearly.
        }
        \label{fig:figV_0}
    \end{minipage}%
    \hfill
    \begin{minipage}[t]{0.49\linewidth}
        \vspace{0pt}
        \centering
        \includegraphics[width=\linewidth]{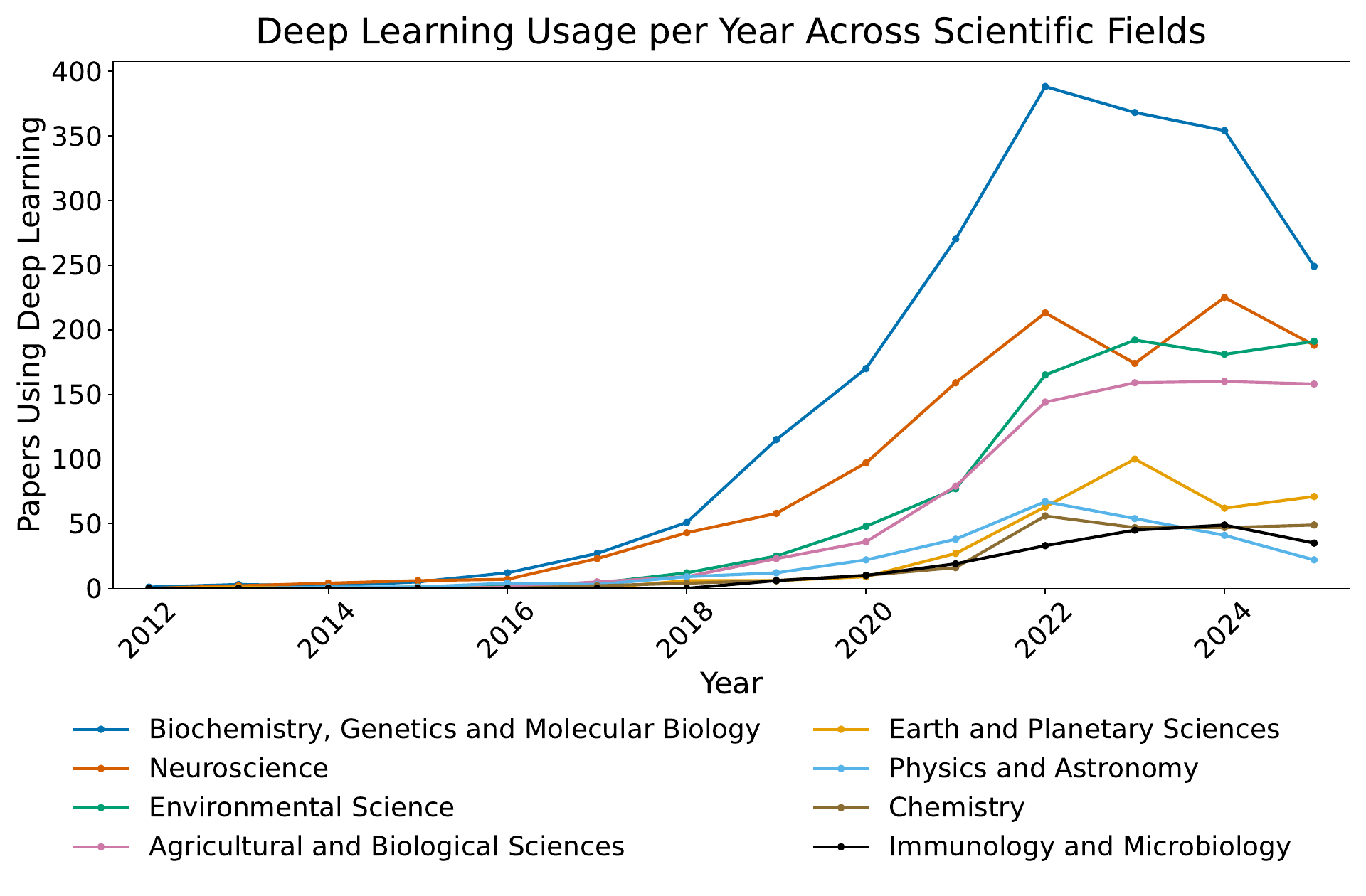}

        \captionof{figure}{
            Annual number of papers using deep learning, by natural science field, from 2012 to 2025. 
            Field counts are based on unique papers assigned to each field by OpenAlex. 
        }
        \label{fig:figV_1}
    \end{minipage}
    \vspace{-15pt}
\end{figure*}
\subsubsection{Deep Learning Usage}
Using Prompt 1, we identified papers employing DL methods within our candidate set. 
Of the~\markdownPapers candidates,~\papersUsingDeepLearning (\papersUsingDeepLearningProp) utilized DL methods. 
The earliest identified DL paper was published in 2009~\citep{10.1371/journal.pone.0006393}, though consistent annual growth began in 2012~\citep{qi_unified_2012}. 
However, non-DL papers still outnumber DL papers annually (\Cref{fig:figV_0}).

When aggregating publications by unique field\footnote{Because OpenAlex may assign multiple fields to a single paper, papers can be included in more than one field total.
}, the three fields with the largest volume of DL using publications are
    \openalexDLFieldOne (\openalexDLFieldOneCount papers), 
    \openalexDLFieldTwo (\openalexDLFieldTwoCount papers), and 
    \openalexDLFieldThree (\openalexDLFieldThreeCount papers). 
\openalexDLFieldOne represents both the largest cumulative and annual growth (\Cref{fig:figV_1}).

\begin{figure*}[htbb]
    \centering

    \begin{minipage}[t]{0.49\linewidth}
        \vspace{0pt}
        \centering
        \includegraphics[width=\linewidth]{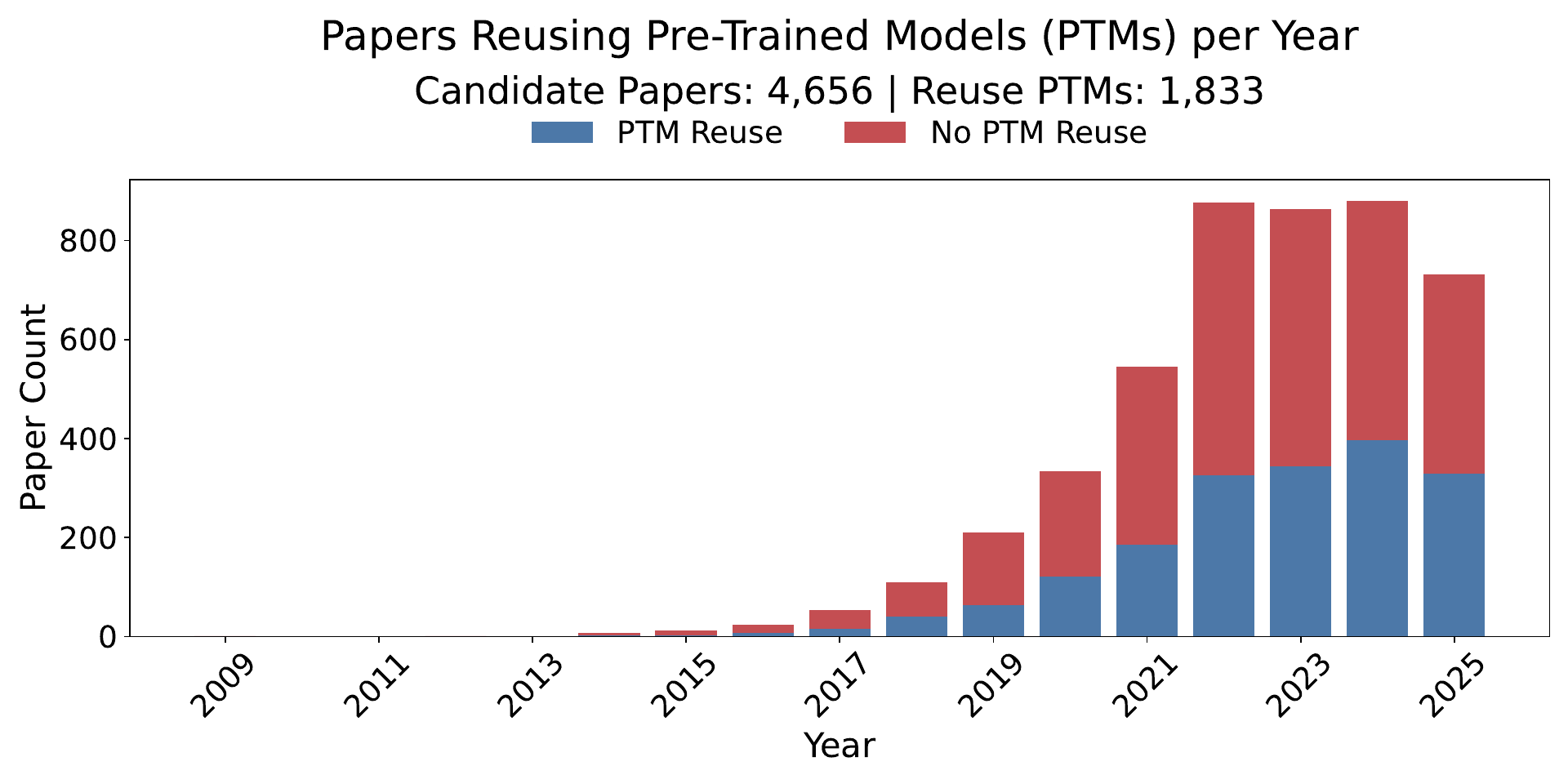}

        \captionof{figure}{
            Number of papers identified reusing PTMs per year plotted against
            papers that do not reuse PTMs. 
            The earliest identified paper that
            reuses PTMs was published in
            2009~\citep{10.1371/journal.pone.0006393},
            however it was not
            until 
            2016~\citep{ptms2}
            that the number of papers
            reusing PTMs began increasing yearly.            
        }
        \label{fig:figW_0}
    \end{minipage}%
    \hfill
    \begin{minipage}[t]{0.49\linewidth}
        \vspace{0pt}
        \centering
        \includegraphics[width=\linewidth]{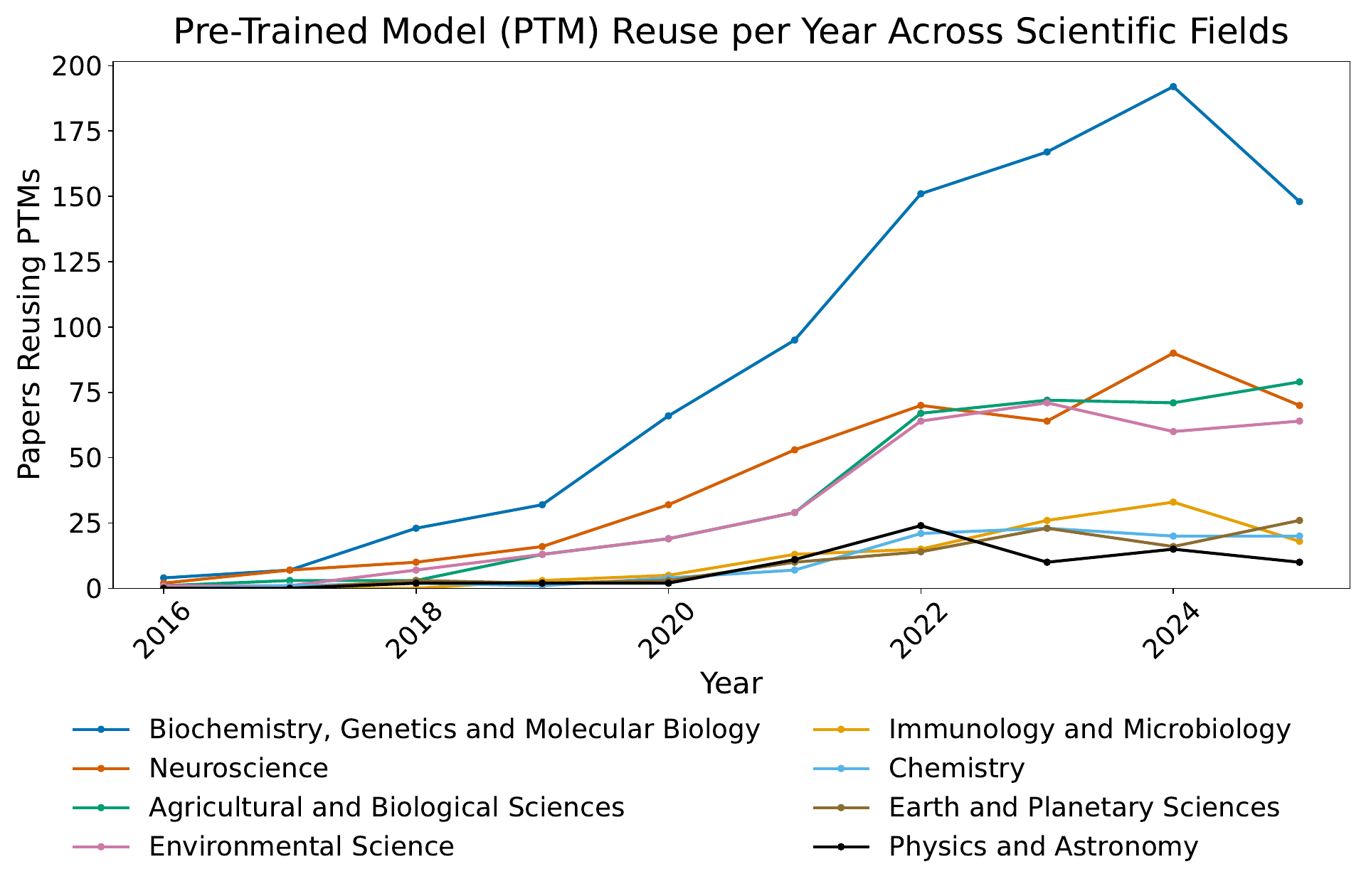}

        \captionof{figure}{%
            Annual number of papers reusing PTMs, by natural science field, from 2012 to 2025. 
            Field counts are based on unique papers assigned to each field by OpenAlex.
        }
        \label{fig:figW_1}
    \end{minipage}

    \vspace{-15pt}
\end{figure*}
\subsubsection{Pre-Trained Model Reuse}
Using Prompt 2, we identified papers reusing PTMs within our candidate set of DL using natural science publications. 
Of the~\papersUsingDeepLearning candidates,~\papersResuingPTMs (\papersResuingPTMsProp) reused PTMs.
The earliest identified PTM reusing paper was published in 2009~\citep{10.1371/journal.pone.0006393}, though consistent annual growth began in 2016~\citep{ptms2}.
Similar to DL usage, non-PTM reusing papers outnumber PTM reusing papers annually (\Cref{fig:figW_0}).

When aggregating publications by unique field, the three fields with the largest volume of PTM reusing publications are 
    \openalexPTMFieldOne (\openalexPTMFieldOneCount papers), 
    \openalexPTMFieldTwo (\openalexPTMFieldTwoCount papers), and 
    \openalexPTMFieldThree (\openalexPTMFieldThreeCount papers). 
\openalexPTMFieldOne represents both the largest cumulative and annual growth (\Cref{fig:figW_1}).

\begin{figure*}[h]
    \centering
    \includegraphics[width=\linewidth]{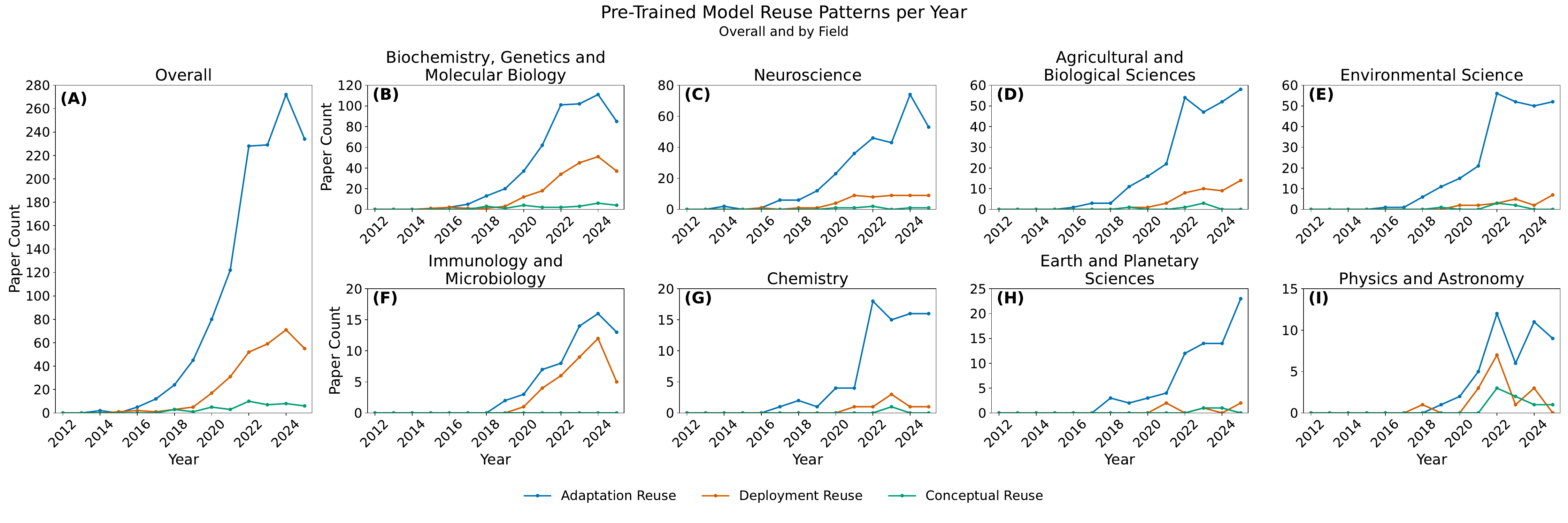}

    \captionof{figure}{
    Counts of PTM reuse patterns per year, 2012–2025. Adaptation reuse is the most common pattern and grows steadily over time, while deployment reuse appears less frequently and conceptual reuse remains relatively rare.
    }
    \label{fig:figXT}
    \vspace{-15pt}
\end{figure*}
\subsubsection{Pre-Trained Model Reuse Pattern Utilization}\label{RQ1_Specific-PTMs}
Using Prompt 4, we identified PTM reuse patterns within our candidate set of PTM reusing natural science publications. 
Of the~\papersResuingPTMs papers reusing PTMs, \ptmAdaptationReuse (\ptmAdaptationReuseProp) leveraged ``adaptation,'' \ptmConceptualReuse (\ptmConceptualReuseProp) leveraged ``conceptual,'' and \ptmDeploymentReuse (\ptmDeploymentReuseProp) leveraged ``deployment'' reuse patterns.

The earliest identified ``adaptation'' reuse was in 2009~\citep{10.1371/journal.pone.0006393}, ``conceptual'' in 2018~\citep{chen_lailaps-qsm_2018}, and ``deployment'' in 2015~\citep{bach_pixel-wise_2015}. ``Adaptation'' has seen the largest usage, followed by ``deployment'' and ``conceptual'' (\Cref{fig:figXT}).

When aggregated by field and reuse pattern (\Cref{fig:figXT}), ``Biochemistry, Genetics and Molecular Biology,'' ``Neuroscience,'' and ``Agricultural and Biological Sciences'' emerge as primary drivers of ``adaptation'' and ``deployment'' reuse, while ``Environmental Science,'' ``Neuroscience,'' and ``Physics and Astronomy'' leverage ``conceptual'' reuse more than ``Agricultural and Biological Sciences.'' Longitudinal data shows all natural science fields have increasingly leveraged the ``adaptation'' reuse pattern.

\section{RQ2: PTM Integration In Scientific Workflow}\label{sec:rq2}
This section details our methodology for identifying what stages of the scientific workflow are being augmented via pre-trained deep learning model (PTM) reuse within natural science between~\analysisStartDate and~\analysisEndDate.

\subsection{Methods}
To identify how PTM reuse has augmented the scientific workflow within natural sciences, we developed a pipeline designed to transition from specific identification of PTMs to the greater context in which they are reused.
First, we isolate sections of the prose specifically mentioning PTMs as identified from our previous analysis (\Cref{sec:rq2-methods-isolation}).
We subsequently employ a large language model (LLM) to map these isolated prose segments against a formalized taxonomy of the scientific workflow (\Cref{sec:rq2-methods-classification}). 
This allows us to categorize the specific functional role of each PTM utilization instance and determine which scientific stage is being augmented by model reuse.
We report the methodology for each step in this section.

\subsubsection{Isolation of Pre-Trained Model Reuse Prose}\label{sec:rq2-methods-isolation}
To identify the PTM reuse augmented step of the scientific process, we first identify sections of a paper's prose that mention previously identified PTMs (\Cref{RQ1_Specific-PTMs}). 
While our previous methods returned the most salient prose to confirm PTM names and reuse patterns, they did not return the section location. 
To remedy this, we parse the Markdown document for the PTM name and return the section content referencing the model.

\subsubsection{Classification of Impacted Scientific Process}\label{sec:rq2-methods-classification}
To classify the steps of the scientific process impacted by PTM reuse, we leverage \verb|gpt-5.4-nano-2026-03-17| with the same parameters outlined in~\Cref{sec:rq1-methods-load}. 
As there is limited consensus on a definitive scientific method~\citep{hepburn_scientific_2026, weinberg_methods_1995}, we label prose under the hypothetico-deductive (H-D) framework~\citep{hempel_philosophy_1966}: 
    ``Observation'' of a natural phenomenon or gap in empirical data requiring analysis,
    ``Hypothesis'' where a PTM's learned representations are proposed to explain the observation,
    ``Background'' where the PTM derives testable predictions about unobserved natural states or experimental outcomes,
    ``Test'' using a PTM to inference on experimental data to derive new insights, and
    ``Analysis'' of whether the PTM's outputs corroborate scientific theory or validly represent the natural system. 
We map this framework to common paper sections: 
    ``Introduction'', 
    ``Methodology'', 
    ``Experimental Design'', 
    ``Results'', and 
    ``Discussion'' respectively.

\subsection{Data Analysis}\label{sec:rq2-analysis}
To identify how PTM reuse patterns have impacted natural science, we analyzed trends in publications reusing PTMs overall, annually, by discipline, and at the intersection of field and year.

\begin{figure}[h]
    \centering
    \includegraphics[width=\linewidth]{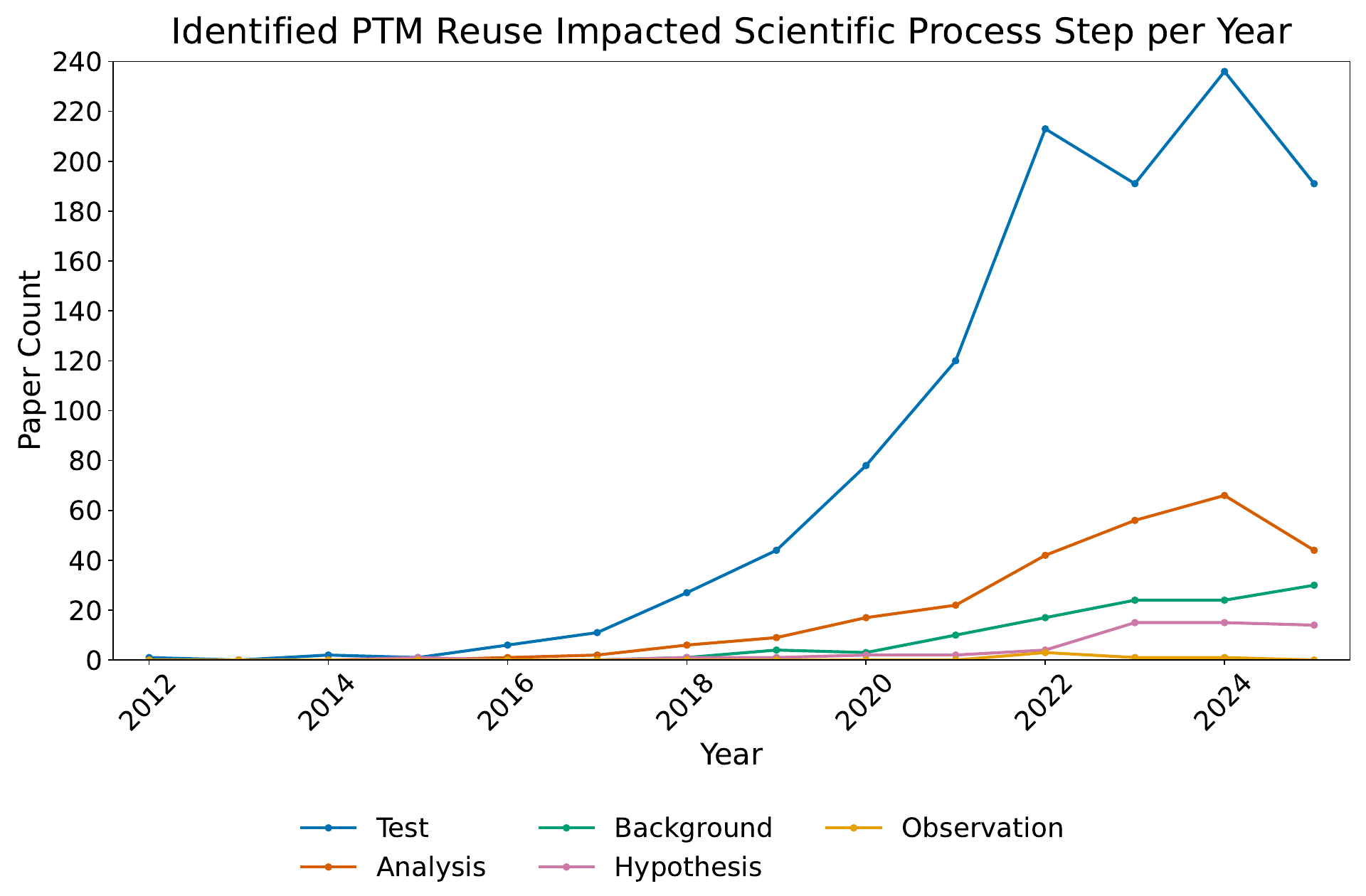}
    \caption{
    Scientific process stages affected by PTM reuse from 2012 to 2025. 
    Annual counts show PTM reuse across five stages of the scientific process: 
        observation, 
        background, 
        hypothesis, 
        analysis, and 
        testing. 
    Reuse occurs most frequently during testing, while the other stages show comparatively fewer instances.
    }
    \label{fig:figU}
    \vspace{-15pt}
\end{figure}
\subsubsection{PTM Reuse Prose Classification}
Using Prompt 5, we identify the specific stage of the scientific workflow where PTMs have made an impact within our candidate set. 
Of the~\papersResuingPTMs papers reusing PTMs, 
    ~\impactTest (\impactTestProp) leveraged PTMs in the ``Test'' stage, 
    ~\impactAnalysis (\impactAnalysisProp) in ``Analysis,'' 
    ~\impactBackground (\impactBackgroundProp) in ``Background,'' 
    ~\impactObservation (\impactObservationProp) in ``Observation,'' and 
    ~\impactHypothesis (\impactHypothesisProp) in ``Hypothesis.''

The earliest identified papers per stage were: 
    ``Test'' in 2009~\citep{10.1371/journal.pone.0006393}, 
    ``Hypothesis'' in 2015~\citep{10.1371/journal.pone.0141287}, 
    ``Analysis'' in 2016~\citep{10.1371/journal.pcbi.1004962}, 
    ``Background'' in 2018~\citep{10.3389/fninf.2018.00023}, and 
    ``Observation'' in 2022~\citep{10.1371/journal.pcbi.1010221}. 
The ``Test'' stage has seen the most PTM impact by volume, followed by ``Analysis'' and ``Background'' (\Cref{fig:figU}).

\section{Discussion}\label{sec:discussion} 
This study examines how natural scientists integrate PTMs into their workflows for data driven discovery. 
We present a qualitative analysis of~\totalPapersFound peer reviewed, open access papers published in 
    BMJ, 
    F1000, 
    FrontiersIn, and 
    PLOS 
from~\analysisStartDate to~\analysisEndDate, serving as a proxy for evaluating the current landscape of PTM reuse across the scientific community.

\subsection{Implications For Science}
\myparagraph{PTM Supply Chain Enabling~\onePTMReusePattern Reuse} 
The findings in~\Cref{sec:rq1} identify a visible increase in deep learning and PTM reuse 
    --- particularly within~\openalexDLFieldOne (\Cref{fig:figW_1}) --- 
and ~\onePTMReusePatternLower PTM reuse is more frequently leveraged than ``deployment'' or ``conceptual'' PTM reuse (\Cref{fig:figW_0}). 
We posit that scientists leverage \onePTMReusePatternLower PTM reuse to realize the benefits of deep learning without incurring the prohibitive costs of training models from scratch, with PTMs serving as reusable scientific infrastructure that shifts the paradigm towards greater integration of existing PTM supply chains into natural science.

\myparagraph{Limited PTM Integration in Scientific Workflow} 
As detailed in~\Cref{sec:rq2}, PTM integration is concentrated within the ``Test'' stage of the scientific workflow (\Cref{fig:figU}).
This suggests that PTMs are currently utilized as evaluative or benchmarking instruments rather than foundational components of early stage experimental design. 
The limited reuse in other stages implies persistent integration barriers\footnote{From anecdotal conversations with scientists, scientists are hesitant to reuse them due to their inherit biases from training.}, 
highlighting a potential missed opportunity to leverage PTMs earlier in the scientific workflow\footnote{It is also possible that PTMs are used elsewhere in the workflow but not disclosed, \eg through undocumented use of LLMs in study design.}.

\subsection{Future Directions}
Future work will 
    curate a human validated dataset mapping papers to PTM reuse patterns via our pipeline and
    investigate the robustness of LLMs for PTM identification through ablation studies. 
We plan to expand this analysis to a broader range of scientific disciplines to determine if common PTM reuse patterns emerge across fields.
Next, we aim to investigate the technical implementation of PTMs within scientific software by analyzing how researchers integrate PTMs into their code
to help establish frameworks for reproducible model reuse for science.

\section{Threats To Validity}\label{sec:threats} 
Several threats to validity may affect 
    our measurements, 
    causal relationships, and 
    generalizability.

Regarding \textit{construct validity}, our methodology relies on 
    keyword searches, 
    OpenAlex's deep learning driven topic assignment, and
    large language models.
Keyword searches may inadvertently exclude valid candidate papers.
Misclassifications from OpenAlex's topic assignments may affect paper selection for our study's population.
We leveraged the \verb|gpt-5.4-nano-2026-03-17| large language model 
    --- due to its competitive performance relative to open models and its more recent training data cutoff ---
to operationalize these constructs, which entail hallucination risks and classification errors.
Reporting bias and varied methodological descriptions in academic prose may also cause us to miss PTM and deep learning reuse candidates. 
Mitigations are detailed in~\Cref{sec:rq1-methods-load}. 

\textit{Internal validity} threats include 
    potential reverse causality, where PTM reuse may drive deep learning adoption rather than the inverse. 
Confounding factors 
    --- such as 
        increased GPU availability, 
        lower financial and technical barriers, and 
        maturing library support ---
could also drive this trend.

Lastly, \textit{external validity} is threatened by our journal selection.
While megajournals were selected to cover a broad selection of the natural sciences, publication trends may differ in scientific communities bound by language (\textit{e.g.}, China).
Our population of identified natural science papers skews towards ``Biochemistry, Genetics and Molecular Biology'' as reported in~\Cref{table:tableJ}, potentially limiting generalizability.
However, across global boundaries the number of published biology and biomedical works have exceeded all other scientific fields~\citep{statisticsncsesPublicationsOutputUS2023}.

\section{Conclusion}
To our knowledge, this is the first empirical study of pre-trained deep learning model (PTM) reuse patterns within the natural sciences. 
By examining how scientists reuse PTMs and report their reuse, we characterize the impact of PTM reuse patterns on the scientific workflow. 
Our findings highlight 
    increasing PTM reuse, 
    dominance of the~\onePTMReusePatternLower pattern, and 
    primary use within the~\topImpactOne stage, 
collectively indicating a maturing pattern of PTM integration characterized by adaptation focused reuse and downstream evaluative application.

\section*{Acknowledgments}
We thank Kimberly Hung for contributions to the visual design of the paper and figures.
This research used resources of the Argonne
Leadership Computing Facility, which is a DOE Office of Science
User Facility supported under Contract DE-AC02-06CH11357.
Davis acknowledges support under Grant No. 2537308, and
Thiruvathukal acknowledges support from the National Science Foundation under Grant No. 2537309.

\balance
\newpage
\bibliographystyle{IEEEtran}

\begin{thebibliography}{100}
\providecommand{\url}[1]{#1}
\csname url@samestyle\endcsname
\providecommand{\newblock}{\relax}
\providecommand{\bibinfo}[2]{#2}
\providecommand{\BIBentrySTDinterwordspacing}{\spaceskip=0pt\relax}
\providecommand{\BIBentryALTinterwordstretchfactor}{4}
\providecommand{\BIBentryALTinterwordspacing}{\spaceskip=\fontdimen2\font plus
\BIBentryALTinterwordstretchfactor\fontdimen3\font minus \fontdimen4\font\relax}
\providecommand{\BIBforeignlanguage}[2]{{%
\expandafter\ifx\csname l@#1\endcsname\relax
\typeout{** WARNING: IEEEtran.bst: No hyphenation pattern has been}%
\typeout{** loaded for the language `#1'. Using the pattern for}%
\typeout{** the default language instead.}%
\else
\language=\csname l@#1\endcsname
\fi
#2}}
\providecommand{\BIBdecl}{\relax}
\BIBdecl

\bibitem{the_royal_swedish_academy_of_sciences_nobel_physics_2024}
\BIBentryALTinterwordspacing
{The Royal Swedish Academy of Sciences}. The nobel prize in physics 2024. [Online]. Available: \url{https://www.nobelprize.org/prizes/physics/2024/press-release/}
\BIBentrySTDinterwordspacing

\bibitem{the_royal_swedish_academy_of_sciences_nobel_chemistry_2024}
\BIBentryALTinterwordspacing
------. The nobel prize in chemistry 2024. [Online]. Available: \url{https://www.nobelprize.org/prizes/chemistry/2024/press-release/}
\BIBentrySTDinterwordspacing

\bibitem{hopfield_neural_1982}
\BIBentryALTinterwordspacing
J.~J. Hopfield, ``Neural networks and physical systems with emergent collective computational abilities.'' \emph{Proceedings of the National Academy of Sciences}, vol.~79, no.~8, pp. 2554--2558, Apr. 1982, publisher: Proceedings of the National Academy of Sciences. [Online]. Available: \url{https://www.pnas.org/doi/abs/10.1073/pnas.79.8.2554}
\BIBentrySTDinterwordspacing

\bibitem{jumper_highly_2021}
J.~Jumper, R.~Evans, A.~Pritzel, T.~Green, M.~Figurnov, O.~Ronneberger, K.~Tunyasuvunakool, R.~Bates, A.~Žídek, A.~Potapenko, A.~Bridgland, C.~Meyer, S.~A.~A. Kohl, A.~J. Ballard, A.~Cowie, B.~Romera-Paredes, S.~Nikolov, R.~Jain, J.~Adler, T.~Back, S.~Petersen, D.~Reiman, E.~Clancy, M.~Zielinski, M.~Steinegger, M.~Pacholska, T.~Berghammer, S.~Bodenstein, D.~Silver, O.~Vinyals, A.~W. Senior, K.~Kavukcuoglu, P.~Kohli, and D.~Hassabis, ``\BIBforeignlanguage{eng}{Highly accurate protein structure prediction with {AlphaFold}},'' \emph{\BIBforeignlanguage{eng}{Nature}}, vol. 596, no. 7873, pp. 583--589, Aug. 2021.

\bibitem{lecun_deep_2015}
\BIBentryALTinterwordspacing
Y.~LeCun, Y.~Bengio, and G.~Hinton, ``\BIBforeignlanguage{en}{Deep learning},'' \emph{\BIBforeignlanguage{en}{Nature}}, vol. 521, no. 7553, pp. 436--444, May 2015, publisher: Nature Publishing Group. [Online]. Available: \url{https://www.nature.com/articles/nature14539}
\BIBentrySTDinterwordspacing

\bibitem{abramson_accurate_2024}
\BIBentryALTinterwordspacing
J.~Abramson, J.~Adler, J.~Dunger, R.~Evans, T.~Green, A.~Pritzel, O.~Ronneberger, L.~Willmore, A.~J. Ballard, J.~Bambrick, S.~W. Bodenstein, D.~A. Evans, C.-C. Hung, M.~O’Neill, D.~Reiman, K.~Tunyasuvunakool, Z.~Wu, A.~Žemgulytė, E.~Arvaniti, C.~Beattie, O.~Bertolli, A.~Bridgland, A.~Cherepanov, M.~Congreve, A.~I. Cowen-Rivers, A.~Cowie, M.~Figurnov, F.~B. Fuchs, H.~Gladman, R.~Jain, Y.~A. Khan, C.~M.~R. Low, K.~Perlin, A.~Potapenko, P.~Savy, S.~Singh, A.~Stecula, A.~Thillaisundaram, C.~Tong, S.~Yakneen, E.~D. Zhong, M.~Zielinski, A.~Žídek, V.~Bapst, P.~Kohli, M.~Jaderberg, D.~Hassabis, and J.~M. Jumper, ``\BIBforeignlanguage{en}{Accurate structure prediction of biomolecular interactions with {AlphaFold} 3},'' \emph{\BIBforeignlanguage{en}{Nature}}, vol. 630, no. 8016, pp. 493--500, Jun. 2024, publisher: Nature Publishing Group. [Online]. Available: \url{https://www.nature.com/articles/s41586-024-07487-w}
\BIBentrySTDinterwordspacing

\bibitem{han_pre-trained_2021}
\BIBentryALTinterwordspacing
X.~Han, Z.~Zhang, N.~Ding, Y.~Gu, X.~Liu, Y.~Huo, J.~Qiu, Y.~Yao, A.~Zhang, L.~Zhang, W.~Han, M.~Huang, Q.~Jin, Y.~Lan, Y.~Liu, Z.~Liu, Z.~Lu, X.~Qiu, R.~Song, J.~Tang, J.-R. Wen, J.~Yuan, W.~X. Zhao, and J.~Zhu, ``Pre-trained models: {Past}, present and future,'' \emph{AI Open}, vol.~2, pp. 225--250, Jan. 2021. [Online]. Available: \url{https://www.sciencedirect.com/science/article/pii/S2666651021000231}
\BIBentrySTDinterwordspacing

\bibitem{davis_reusing_2023}
\BIBentryALTinterwordspacing
J.~C. Davis, P.~Jajal, W.~Jiang, T.~R. Schorlemmer, N.~Synovic, and G.~K. Thiruvathukal, ``Reusing {Deep} {Learning} {Models}: {Challenges} and {Directions} in {Software} {Engineering},'' in \emph{2023 {IEEE} {John} {Vincent} {Atanasoff} {International} {Symposium} on {Modern} {Computing} ({JVA})}, Jul. 2023, pp. 17--30. [Online]. Available: \url{https://ieeexplore.ieee.org/abstract/document/10387479}
\BIBentrySTDinterwordspacing

\bibitem{jiang_empirical_2023}
\BIBentryALTinterwordspacing
W.~Jiang, N.~Synovic, M.~Hyatt, T.~R. Schorlemmer, R.~Sethi, Y.-H. Lu, G.~K. Thiruvathukal, and J.~C. Davis, ``An {Empirical} {Study} of {Pre}-{Trained} {Model} {Reuse} in the {Hugging} {Face} {Deep} {Learning} {Model} {Registry},'' in \emph{Proceedings of the 45th {International} {Conference} on {Software} {Engineering}}, ser. {ICSE} '23.\hskip 1em plus 0.5em minus 0.4em\relax Melbourne, Victoria, Australia: IEEE Press, Jul. 2023, pp. 2463--2475. [Online]. Available: \url{https://dl.acm.org/doi/10.1109/ICSE48619.2023.00206}
\BIBentrySTDinterwordspacing

\bibitem{tang_empirical_2021}
\BIBentryALTinterwordspacing
Y.~Tang, R.~Khatchadourian, M.~Bagherzadeh, R.~Singh, A.~Stewart, and A.~Raja, ``An empirical study of refactorings and technical debt in machine learning systems,'' in \emph{Proceedings of the 43rd International Conference on Software Engineering}, ser. {ICSE} '21.\hskip 1em plus 0.5em minus 0.4em\relax {IEEE} Press, pp. 238--250. [Online]. Available: \url{https://dl.acm.org/doi/10.1109/ICSE43902.2021.00033}
\BIBentrySTDinterwordspacing

\bibitem{sculley_hidden_2015}
\BIBentryALTinterwordspacing
D.~Sculley, G.~Holt, D.~Golovin, E.~Davydov, T.~Phillips, D.~Ebner, V.~Chaudhary, M.~Young, J.-F. Crespo, and D.~Dennison, ``Hidden {Technical} {Debt} in {Machine} {Learning} {Systems},'' in \emph{Advances in {Neural} {Information} {Processing} {Systems}}, vol.~28.\hskip 1em plus 0.5em minus 0.4em\relax Curran Associates, Inc., 2015. [Online]. Available: \url{https://proceedings.neurips.cc/paper_files/paper/2015/hash/86df7dcfd896fcaf2674f757a2463eba-Abstract.html}
\BIBentrySTDinterwordspacing

\bibitem{menshawy_navigating_2024}
\BIBentryALTinterwordspacing
A.~Menshawy, Z.~Nawaz, and M.~Fahmy, ``Navigating challenges and technical debt in large language models deployment,'' in \emph{Proceedings of the 4th Workshop on Machine Learning and Systems}, ser. {EuroMLSys} '24.\hskip 1em plus 0.5em minus 0.4em\relax Association for Computing Machinery, pp. 192--199. [Online]. Available: \url{https://doi.org/10.1145/3642970.3655840}
\BIBentrySTDinterwordspacing

\bibitem{carver_software_2016}
J.~C. Carver, N.~P.~C. Hong, and G.~K. Thiruvathukal, \emph{\BIBforeignlanguage{en}{Software {Engineering} for {Science}}}.\hskip 1em plus 0.5em minus 0.4em\relax CRC Press, Nov. 2016, google-Books-ID: xSgNDgAAQBAJ.

\bibitem{ching_opportunities_2018}
\BIBentryALTinterwordspacing
T.~Ching, D.~S. Himmelstein, B.~K. Beaulieu-Jones, A.~A. Kalinin, B.~T. Do, G.~P. Way, E.~Ferrero, P.-M. Agapow, M.~Zietz, M.~M. Hoffman, W.~Xie, G.~L. Rosen, B.~J. Lengerich, J.~Israeli, J.~Lanchantin, S.~Woloszynek, A.~E. Carpenter, A.~Shrikumar, J.~Xu, E.~M. Cofer, C.~A. Lavender, S.~C. Turaga, A.~M. Alexandari, Z.~Lu, D.~J. Harris, D.~{DeCaprio}, Y.~Qi, A.~Kundaje, Y.~Peng, L.~K. Wiley, M.~H.~S. Segler, S.~M. Boca, S.~J. Swamidass, A.~Huang, A.~Gitter, and C.~S. Greene, ``Opportunities and obstacles for deep learning in biology and medicine,'' pages: 142760 Section: New Results. [Online]. Available: \url{https://www.biorxiv.org/content/10.1101/142760v2}
\BIBentrySTDinterwordspacing

\bibitem{mater_deep_2019}
\BIBentryALTinterwordspacing
A.~C. Mater and M.~L. Coote, ``Deep learning in chemistry,'' vol.~59, no.~6, pp. 2545--2559, publisher: American Chemical Society. [Online]. Available: \url{https://doi.org/10.1021/acs.jcim.9b00266}
\BIBentrySTDinterwordspacing

\bibitem{tanaka_deep_2021}
\BIBentryALTinterwordspacing
A.~Tanaka, A.~Tomiya, and K.~Hashimoto, \emph{Deep Learning and Physics}, ser. Mathematical Physics Studies.\hskip 1em plus 0.5em minus 0.4em\relax Springer. [Online]. Available: \url{http://link.springer.com/10.1007/978-981-33-6108-9}
\BIBentrySTDinterwordspacing

\bibitem{bjork_evolution_2018}
\BIBentryALTinterwordspacing
B.-C. Björk, ``Evolution of the scholarly mega-journal, 2006–2017,'' vol.~6, p. e4357, publisher: {PeerJ} Inc. [Online]. Available: \url{https://peerj.com/articles/4357}
\BIBentrySTDinterwordspacing

\bibitem{synovicEmpiricalInvestigationPreTrained2026aZenodo}
\BIBentryALTinterwordspacing
N.~Synovic, K.~Ryzka, A.~V. Vellucci~Solari, K.~Lyons, J.~C. Davis, and G.~K. Thiruvathukal, ``An empirical investigation of pre-trained deep learning model reuse in the scientific process artifact.'' [Online]. Available: \url{https://zenodo.org/records/21842942}
\BIBentrySTDinterwordspacing

\bibitem{lachance_practical_2020}
\BIBentryALTinterwordspacing
J.~{LaChance} and D.~J. Cohen, ``Practical fluorescence reconstruction microscopy for large samples and low-magnification imaging,'' vol.~16, no.~12, p. e1008443. [Online]. Available: \url{https://journals.plos.org/ploscompbiol/article?id=10.1371/journal.pcbi.1008443}
\BIBentrySTDinterwordspacing

\bibitem{pan_generalizing_2023}
\BIBentryALTinterwordspacing
X.~Pan, A.~{DeForge}, and O.~Schwartz, ``Generalizing biological surround suppression based on center surround similarity via deep neural network models,'' vol.~19, no.~9, p. e1011486. [Online]. Available: \url{https://journals.plos.org/ploscompbiol/article?id=10.1371/journal.pcbi.1011486}
\BIBentrySTDinterwordspacing

\bibitem{dutagaci_characterization_2023}
\BIBentryALTinterwordspacing
B.~Dutagaci, B.~Duan, C.~Qiu, C.~D. Kaplan, and M.~Feig, ``Characterization of {RNA} polymerase {II} trigger loop mutations using molecular dynamics simulations and machine learning,'' vol.~19, no.~3, p. e1010999. [Online]. Available: \url{https://journals.plos.org/ploscompbiol/article?id=10.1371/journal.pcbi.1010999}
\BIBentrySTDinterwordspacing

\bibitem{rattray_machine_2023}
\BIBentryALTinterwordspacing
J.~B. Rattray, R.~J. Lowhorn, R.~Walden, P.~Márquez-Zacarías, E.~Molotkova, G.~Perron, C.~Solis-Lemus, D.~P. Alarcon, and S.~P. Brown, ``Machine learning identification of pseudomonas aeruginosa strains from colony image data,'' vol.~19, no.~12, p. e1011699. [Online]. Available: \url{https://journals.plos.org/ploscompbiol/article?id=10.1371/journal.pcbi.1011699}
\BIBentrySTDinterwordspacing

\bibitem{ha_classification_2023}
\BIBentryALTinterwordspacing
Y.~Ha and S.-S. Kim, ``Classification of large ornithopod dinosaur footprints using xception transfer learning,'' vol.~18, no.~11, p. e0293020. [Online]. Available: \url{https://journals.plos.org/plosone/article?id=10.1371/journal.pone.0293020}
\BIBentrySTDinterwordspacing

\bibitem{cai_detection_2021}
\BIBentryALTinterwordspacing
Y.~Cai, X.~Zhang, S.~Z. Kovalsky, H.~T. Ghashghaei, and A.~Greenbaum, ``Detection and classification of neurons and glial cells in the {MADM} mouse brain using {RetinaNet},'' vol.~16, no.~9, p. e0257426. [Online]. Available: \url{https://journals.plos.org/plosone/article?id=10.1371/journal.pone.0257426}
\BIBentrySTDinterwordspacing

\bibitem{dick_differential_2020}
\BIBentryALTinterwordspacing
F.~Dick, G.~S. Nido, G.~W. Alves, O.-B. Tysnes, G.~H. Nilsen, C.~Dölle, and C.~Tzoulis, ``Differential transcript usage in the parkinson’s disease brain,'' vol.~16, no.~11, p. e1009182. [Online]. Available: \url{https://journals.plos.org/plosgenetics/article?id=10.1371/journal.pgen.1009182}
\BIBentrySTDinterwordspacing

\bibitem{oyarbide_sperm_2023}
\BIBentryALTinterwordspacing
U.~Oyarbide, L.~J. Feyrer, and J.~Gordon, ``Sperm and northern bottlenose whale interactions with deep-water trawlers in the western north atlantic,'' vol.~18, no.~8, p. e0289626. [Online]. Available: \url{https://journals.plos.org/plosone/article?id=10.1371/journal.pone.0289626}
\BIBentrySTDinterwordspacing

\bibitem{mathew_genome_2019}
\BIBentryALTinterwordspacing
R.~Mathew and C.~H. Opperman, ``The genome of the migratory nematode, radopholus similis, reveals signatures of close association to the sedentary cyst nematodes,'' vol.~14, no.~10, p. e0224391. [Online]. Available: \url{https://journals.plos.org/plosone/article?id=10.1371/journal.pone.0224391}
\BIBentrySTDinterwordspacing

\bibitem{jiang_peatmoss_2024}
\BIBentryALTinterwordspacing
W.~Jiang, J.~Yasmin, J.~Jones, N.~Synovic, J.~Kuo, N.~Bielanski, Y.~Tian, G.~K. Thiruvathukal, and J.~C. Davis, ``{PeaTMOSS}: A dataset and initial analysis of pre-trained models in open-source software,'' in \emph{Proceedings of the 21st International Conference on Mining Software Repositories}, ser. {MSR} '24.\hskip 1em plus 0.5em minus 0.4em\relax Association for Computing Machinery, pp. 431--443. [Online]. Available: \url{https://dl.acm.org/doi/10.1145/3643991.3644907}
\BIBentrySTDinterwordspacing

\bibitem{jiang_empirical_2022}
\BIBentryALTinterwordspacing
W.~Jiang, N.~Synovic, R.~Sethi, A.~Indarapu, M.~Hyatt, T.~R. Schorlemmer, G.~K. Thiruvathukal, and J.~C. Davis, ``An {Empirical} {Study} of {Artifacts} and {Security} {Risks} in the {Pre}-trained {Model} {Supply} {Chain},'' in \emph{Proceedings of the 2022 {ACM} {Workshop} on {Software} {Supply} {Chain} {Offensive} {Research} and {Ecosystem} {Defenses}}, ser. {SCORED}'22.\hskip 1em plus 0.5em minus 0.4em\relax New York, NY, USA: Association for Computing Machinery, Nov. 2022, pp. 105--114. [Online]. Available: \url{https://dl.acm.org/doi/10.1145/3560835.3564547}
\BIBentrySTDinterwordspacing

\bibitem{schwartz_green_2020}
\BIBentryALTinterwordspacing
R.~Schwartz, J.~Dodge, N.~A. Smith, and O.~Etzioni, ``Green {AI},'' \emph{Commun. ACM}, vol.~63, no.~12, pp. 54--63, Nov. 2020. [Online]. Available: \url{https://dl.acm.org/doi/10.1145/3381831}
\BIBentrySTDinterwordspacing

\bibitem{strubell_energy_2020}
\BIBentryALTinterwordspacing
E.~Strubell, A.~Ganesh, and A.~McCallum, ``\BIBforeignlanguage{en}{Energy and {Policy} {Considerations} for {Modern} {Deep} {Learning} {Research}},'' \emph{\BIBforeignlanguage{en}{Proceedings of the AAAI Conference on Artificial Intelligence}}, vol.~34, no.~09, pp. 13\,693--13\,696, Apr. 2020, number: 09. [Online]. Available: \url{https://ojs.aaai.org/index.php/AAAI/article/view/7123}
\BIBentrySTDinterwordspacing

\bibitem{wolf_huggingfaces_2020}
\BIBentryALTinterwordspacing
T.~Wolf, L.~Debut, V.~Sanh, J.~Chaumond, C.~Delangue, A.~Moi, P.~Cistac, T.~Rault, R.~Louf, M.~Funtowicz, J.~Davison, S.~Shleifer, P.~v. Platen, C.~Ma, Y.~Jernite, J.~Plu, C.~Xu, T.~L. Scao, S.~Gugger, M.~Drame, Q.~Lhoest, and A.~M. Rush, ``{HuggingFace}'s {Transformers}: {State}-of-the-art {Natural} {Language} {Processing},'' Jul. 2020, arXiv:1910.03771. [Online]. Available: \url{http://arxiv.org/abs/1910.03771}
\BIBentrySTDinterwordspacing

\bibitem{ansel_pytorch_2024}
\BIBentryALTinterwordspacing
J.~Ansel, E.~Yang, H.~He, N.~Gimelshein, A.~Jain, M.~Voznesensky, B.~Bao, P.~Bell, D.~Berard, E.~Burovski, G.~Chauhan, A.~Chourdia, W.~Constable, A.~Desmaison, Z.~DeVito, E.~Ellison, W.~Feng, J.~Gong, M.~Gschwind, B.~Hirsh, S.~Huang, K.~Kalambarkar, L.~Kirsch, M.~Lazos, M.~Lezcano, Y.~Liang, J.~Liang, Y.~Lu, C.~Luk, B.~Maher, Y.~Pan, C.~Puhrsch, M.~Reso, M.~Saroufim, M.~Y. Siraichi, H.~Suk, M.~Suo, P.~Tillet, E.~Wang, X.~Wang, W.~Wen, S.~Zhang, X.~Zhao, K.~Zhou, R.~Zou, A.~Mathews, G.~Chanan, P.~Wu, and S.~Chintala, ``{PyTorch} 2: {Faster} {Machine} {Learning} {Through} {Dynamic} {Python} {Bytecode} {Transformation} and {Graph} {Compilation},'' in \emph{29th {ACM} {International} {Conference} on {Architectural} {Support} for {Programming} {Languages} and {Operating} {Systems}, {Volume} 2 ({ASPLOS} '24)}.\hskip 1em plus 0.5em minus 0.4em\relax ACM, Apr. 2024. [Online]. Available: \url{https://pytorch.org/assets/pytorch2-2.pdf}
\BIBentrySTDinterwordspacing

\bibitem{prottasha_peft_2025}
\BIBentryALTinterwordspacing
N.~J. Prottasha, U.~R. Chowdhury, S.~Mohanto, T.~Nuzhat, A.~A. Sami, M.~S. Ali, M.~S.~I. Sobuj, H.~Raman, M.~Kowsher, and O.~O. Garibay, ``{PEFT} a2z: Parameter-efficient fine-tuning survey for large language and vision models.'' [Online]. Available: \url{http://arxiv.org/abs/2504.14117}
\BIBentrySTDinterwordspacing

\bibitem{huggingface_2024}
\BIBentryALTinterwordspacing
``Hugging {Face} – {The} {AI} community building the future.'' Nov. 2024. [Online]. Available: \url{https://huggingface.co/}
\BIBentrySTDinterwordspacing

\bibitem{yu_tensorflow_2020}
\BIBentryALTinterwordspacing
H.~Yu, C.~Chen, X.~Du, Y.~Li, A.~Rashwan, L.~Hou, P.~Jin, F.~Yang, F.~Liu, J.~Kim, and J.~Li, ``{TensorFlow} {Model} {Garden},'' 2020. [Online]. Available: \url{https://github.com/tensorflow/models}
\BIBentrySTDinterwordspacing

\bibitem{jiang_challenges_2024}
\BIBentryALTinterwordspacing
W.~Jiang, V.~Banna, N.~Vivek, A.~Goel, N.~Synovic, G.~K. Thiruvathukal, and J.~C. Davis, ``Challenges and practices of deep learning model reengineering: {A} case study on computer vision,'' \emph{Empirical Softw. Engg.}, vol.~29, no.~6, Aug. 2024. [Online]. Available: \url{https://doi.org/10.1007/s10664-024-10521-0}
\BIBentrySTDinterwordspacing

\bibitem{jajal_analysis_2023}
\BIBentryALTinterwordspacing
P.~Jajal, W.~Jiang, A.~Tewari, J.~Woo, G.~K. Thiruvathukal, and J.~C. Davis, ``Analysis of {Failures} and {Risks} in {Deep} {Learning} {Model} {Converters}: {A} {Case} {Study} in the {ONNX} {Ecosystem},'' Oct. 2023, arXiv:2303.17708 version: 2. [Online]. Available: \url{http://arxiv.org/abs/2303.17708}
\BIBentrySTDinterwordspacing

\bibitem{duede_oil_2024}
\BIBentryALTinterwordspacing
E.~Duede, W.~Dolan, A.~Bauer, I.~Foster, and K.~Lakhani, ``Oil \& {Water}? {Diffusion} of {AI} {Within} and {Across} {Scientific} {Fields},'' May 2024, arXiv:2405.15828. [Online]. Available: \url{http://arxiv.org/abs/2405.15828}
\BIBentrySTDinterwordspacing

\bibitem{angermueller_deep_2016}
\BIBentryALTinterwordspacing
C.~Angermueller, T.~Pärnamaa, L.~Parts, and O.~Stegle, ``Deep learning for computational biology,'' vol.~12, no.~7, p. 878, publisher: John Wiley \& Sons, Ltd. [Online]. Available: \url{https://www.embopress.org/doi/full/10.15252/msb.20156651}
\BIBentrySTDinterwordspacing

\bibitem{ramsundar_deep_2019}
B.~Ramsundar, P.~Eastman, P.~Walters, and V.~Pande, \emph{Deep Learning for the Life Sciences: Applying Deep Learning to Genomics, Microscopy, Drug Discovery, and More}.\hskip 1em plus 0.5em minus 0.4em\relax O'Reilly Media.

\bibitem{bergen_machine_2019}
\BIBentryALTinterwordspacing
K.~J. Bergen, P.~A. Johnson, M.~V. de~Hoop, and G.~C. Beroza, ``Machine learning for data-driven discovery in solid earth geoscience,'' vol. 363, no. 6433, p. eaau0323, publisher: American Association for the Advancement of Science. [Online]. Available: \url{https://www.science.org/doi/10.1126/science.aau0323}
\BIBentrySTDinterwordspacing

\bibitem{zambaldi_novo_2024}
\BIBentryALTinterwordspacing
V.~Zambaldi, D.~La, A.~E. Chu, H.~Patani, A.~E. Danson, T.~O.~C. Kwan, T.~Frerix, R.~G. Schneider, D.~Saxton, A.~Thillaisundaram, Z.~Wu, I.~Moraes, O.~Lange, E.~Papa, G.~Stanton, V.~Martin, S.~Singh, L.~H. Wong, R.~Bates, S.~A. Kohl, J.~Abramson, A.~W. Senior, Y.~Alguel, M.~Y. Wu, I.~M. Aspalter, K.~Bentley, D.~L.~V. Bauer, P.~Cherepanov, D.~Hassabis, P.~Kohli, R.~Fergus, and J.~Wang, ``De novo design of high-affinity protein binders with {AlphaProteo},'' Sep. 2024, arXiv:2409.08022. [Online]. Available: \url{http://arxiv.org/abs/2409.08022}
\BIBentrySTDinterwordspacing

\bibitem{krishna_generalized_2024}
\BIBentryALTinterwordspacing
R.~Krishna, J.~Wang, W.~Ahern, P.~Sturmfels, P.~Venkatesh, I.~Kalvet, G.~R. Lee, F.~S. Morey-Burrows, I.~Anishchenko, I.~R. Humphreys, R.~{McHugh}, D.~Vafeados, X.~Li, G.~A. Sutherland, A.~Hitchcock, C.~N. Hunter, A.~Kang, E.~Brackenbrough, A.~K. Bera, M.~Baek, F.~{DiMaio}, and D.~Baker, ``Generalized biomolecular modeling and design with {RoseTTAFold} all-atom,'' vol. 384, no. 6693, p. eadl2528, publisher: American Association for the Advancement of Science. [Online]. Available: \url{https://www.science.org/doi/10.1126/science.adl2528}
\BIBentrySTDinterwordspacing

\bibitem{schwaller_extraction_2021}
\BIBentryALTinterwordspacing
P.~Schwaller, B.~Hoover, J.-L. Reymond, H.~Strobelt, and T.~Laino, ``Extraction of organic chemistry grammar from unsupervised learning of chemical reactions,'' vol.~7, no.~15, p. eabe4166, publisher: American Association for the Advancement of Science. [Online]. Available: \url{https://www.science.org/doi/10.1126/sciadv.abe4166}
\BIBentrySTDinterwordspacing

\bibitem{zhang_large_2025}
\BIBentryALTinterwordspacing
Y.~Zhang, Y.~Han, S.~Chen, R.~Yu, X.~Zhao, X.~Liu, K.~Zeng, M.~Yu, J.~Tian, F.~Zhu, X.~Yang, Y.~Jin, and Y.~Xu, ``Large language models to accelerate organic chemistry synthesis,'' vol.~7, no.~7, pp. 1010--1022, publisher: Nature Publishing Group. [Online]. Available: \url{https://www.nature.com/articles/s42256-025-01066-y}
\BIBentrySTDinterwordspacing

\bibitem{wang_deep_2025}
\BIBentryALTinterwordspacing
Y.~Wang, C.-Y. Lai, D.~J. Prior, and C.~Cowen-Breen, ``Deep learning the flow law of antarctic ice shelves,'' vol. 387, no. 6739, pp. 1219--1224, publisher: American Association for the Advancement of Science. [Online]. Available: \url{https://www.science.org/doi/10.1126/science.adp3300}
\BIBentrySTDinterwordspacing

\bibitem{mousavi_deep-learning_2022}
\BIBentryALTinterwordspacing
S.~M. Mousavi and G.~C. Beroza, ``Deep-learning seismology,'' vol. 377, no. 6607, p. eabm4470, publisher: American Association for the Advancement of Science. [Online]. Available: \url{https://www.science.org/doi/10.1126/science.abm4470}
\BIBentrySTDinterwordspacing

\bibitem{lee_ten_2022}
\BIBentryALTinterwordspacing
B.~D. Lee, A.~Gitter, C.~S. Greene, S.~Raschka, F.~Maguire, A.~J. Titus, M.~D. Kessler, A.~J. Lee, M.~G. Chevrette, P.~A. Stewart, T.~Britto-Borges, E.~M. Cofer, K.-H. Yu, J.~J. Carmona, E.~J. Fertig, A.~A. Kalinin, B.~Signal, B.~J. Lengerich, T.~J.~T. Jr, and S.~M. Boca, ``Ten quick tips for deep learning in biology,'' vol.~18, no.~3, p. e1009803, publisher: Public Library of Science. [Online]. Available: \url{https://journals.plos.org/ploscompbiol/article?id=10.1371/journal.pcbi.1009803}
\BIBentrySTDinterwordspacing

\bibitem{malviya-thakur_scicat_2023}
\BIBentryALTinterwordspacing
A.~Malviya-Thakur, R.~Milewicz, L.~Paganini, A.~S.~I. Mahmoud, and A.~Mockus, ``{SciCat}: {A} {Curated} {Dataset} of {Scientific} {Software} {Repositories},'' Dec. 2023, arXiv:2312.06382 [cs]. [Online]. Available: \url{http://arxiv.org/abs/2312.06382}
\BIBentrySTDinterwordspacing

\bibitem{howison_software_2016}
\BIBentryALTinterwordspacing
J.~Howison and J.~Bullard, ``\BIBforeignlanguage{en}{Software in the scientific literature: {Problems} with seeing, finding, and using software mentioned in the biology literature},'' \emph{\BIBforeignlanguage{en}{Journal of the Association for Information Science and Technology}}, vol.~67, no.~9, pp. 2137--2155, 2016. [Online]. Available: \url{https://onlinelibrary.wiley.com/doi/abs/10.1002/asi.23538}
\BIBentrySTDinterwordspacing

\bibitem{priem_openalex_2022}
\BIBentryALTinterwordspacing
J.~Priem, H.~Piwowar, and R.~Orr, ``{OpenAlex}: {A} fully-open index of scholarly works, authors, venues, institutions, and concepts,'' Jun. 2022, arXiv:2205.01833. [Online]. Available: \url{http://arxiv.org/abs/2205.01833}
\BIBentrySTDinterwordspacing

\bibitem{synovicEmpiricalInvestigationPreTrained2026}
\BIBentryALTinterwordspacing
N.~M. Synovic, K.~Ryzka, A.~V.~V. Solari, K.~Lyons, J.~C. Davis, and G.~K. Thiruvathukal, ``An empirical investigation of pre-trained deep learning model reuse in the scientific process.'' [Online]. Available: \url{http://arxiv.org/abs/2603.13584}
\BIBentrySTDinterwordspacing

\bibitem{noauthor_oadocumentation_2024}
\BIBentryALTinterwordspacing
{No Author}, ``\BIBforeignlanguage{en}{Overview {\textbar} {OpenAlex} technical documentation},'' Jul. 2024. [Online]. Available: \url{https://docs.openalex.org}
\BIBentrySTDinterwordspacing

\bibitem{encyclopaedia_britannica_natural_2025}
{Encyclopædia Britannica}, ``Natural science.''

\bibitem{national_information_standards_organization_jats_standing_committee_ansiniso_2024}
\BIBentryALTinterwordspacing
{National Information Standards Organization JATS Standing Committee}, ``\BIBforeignlanguage{en}{{ANSI}/{NISO} {Z39}.96-2024, {JATS}: {Journal} {Article} {Tag} {Suite}, version 1.4},'' Oct. 2024. [Online]. Available: \url{https://www.niso.org/publications/z3996-2024-jats}
\BIBentrySTDinterwordspacing

\bibitem{MacFarlane_Pandoc}
\BIBentryALTinterwordspacing
J.~MacFarlane, A.~Krewinkel, and J.~Rosenthal, ``{Pandoc}.'' [Online]. Available: \url{https://github.com/jgm/pandoc}
\BIBentrySTDinterwordspacing

\bibitem{teo_how_2023}
\BIBentryALTinterwordspacing
S.~Teo, ``\BIBforeignlanguage{en-US}{How {I} {Won} {Singapore}'s {GPT}-4 {Prompt} {Engineering} {Competition}},'' Dec. 2023. [Online]. Available: \url{https://towardsdatascience.com/how-i-won-singapores-gpt-4-prompt-engineering-competition-34c195a93d41/}
\BIBentrySTDinterwordspacing

\bibitem{wolf-etal-2020-transformers}
\BIBentryALTinterwordspacing
T.~Wolf, L.~Debut, V.~Sanh, J.~Chaumond, C.~Delangue, A.~Moi, P.~Cistac, T.~Rault, R.~Louf, M.~Funtowicz, J.~Davison, S.~Shleifer, P.~von Platen, C.~Ma, Y.~Jernite, J.~Plu, C.~Xu, T.~L. Scao, S.~Gugger, M.~Drame, Q.~Lhoest, and A.~M. Rush, ``Transformers: State-of-the-art natural language processing,'' in \emph{Proceedings of the 2020 Conference on Empirical Methods in Natural Language Processing: System Demonstrations}.\hskip 1em plus 0.5em minus 0.4em\relax Online: Association for Computational Linguistics, Oct. 2020, pp. 38--45. [Online]. Available: \url{https://www.aclweb.org/anthology/2020.emnlp-demos.6}
\BIBentrySTDinterwordspacing

\bibitem{von_Platen_Diffusers_State-of-the-art_diffusion}
\BIBentryALTinterwordspacing
P.~von Platen, S.~Patil, A.~Lozhkov, P.~Cuenca, N.~Lambert, K.~Rasul, M.~Davaadorj, D.~Nair, S.~Paul, S.~Liu, W.~Berman, Y.~Xu, and T.~Wolf, ``{Diffusers: State-of-the-art diffusion models}.'' [Online]. Available: \url{https://github.com/huggingface/diffusers}
\BIBentrySTDinterwordspacing

\bibitem{bio1}
\BIBentryALTinterwordspacing
A.~L. Tyson, C.~V. Rousseau, C.~J. Niedworok, S.~Keshavarzi, C.~Tsitoura, L.~Cossell, M.~Strom, and T.~W. Margrie, ``A deep learning algorithm for 3d cell detection in whole mouse brain image datasets,'' \emph{PLOS Computational Biology}, vol.~17, no.~5, pp. 1--17, 05 2021. [Online]. Available: \url{https://doi.org/10.1371/journal.pcbi.1009074}
\BIBentrySTDinterwordspacing

\bibitem{bio2}
\BIBentryALTinterwordspacing
X.~Niu, K.~Yang, G.~Zhang, Z.~Yang, and X.~Hu, ``A pretraining-retraining strategy of deep learning improves cell-specific enhancer predictions,'' vol.~10. [Online]. Available: \url{https://www.frontiersin.org/journals/genetics/articles/10.3389/fgene.2019.01305/full}
\BIBentrySTDinterwordspacing

\bibitem{bio3}
\BIBentryALTinterwordspacing
A.~Li, Y.~Deng, Y.~Tan, and M.~Chen, ``A transfer learning-based approach for lysine propionylation prediction,'' vol.~12. [Online]. Available: \url{https://www.frontiersin.org/journals/physiology/articles/10.3389/fphys.2021.658633/full}
\BIBentrySTDinterwordspacing

\bibitem{palanichamy_machine_2022}
\BIBentryALTinterwordspacing
N.~Palanichamy, S.-C. Haw, S.~S, R.~Murugan, and K.~Govindasamy, ``Machine learning methods to predict particulate matter.'' [Online]. Available: \url{https://f1000research.com/articles/11-406}
\BIBentrySTDinterwordspacing

\bibitem{li_fish_2025}
\BIBentryALTinterwordspacing
G.~Li, A.~Lian, Z.~Yao, Y.~Hu, G.~Pang, T.~Yuan, Z.~Li, X.~Huang, and G.~Wang, ``Fish keypoint detection for offshore aquaculture: a robust deep learning approach with {PCA}-based shape constraint,'' vol.~12. [Online]. Available: \url{https://www.frontiersin.org/journals/marine-science/articles/10.3389/fmars.2025.1619457/full}
\BIBentrySTDinterwordspacing

\bibitem{guo_integration_2025}
\BIBentryALTinterwordspacing
J.~Guo, X.~Lin, and Y.~Xiao, ``Integration of smart sensors and phytoremediation for real-time pollution monitoring and ecological restoration in agricultural waste management,'' vol.~16. [Online]. Available: \url{https://www.frontiersin.org/journals/plant-science/articles/10.3389/fpls.2025.1550302/full}
\BIBentrySTDinterwordspacing

\bibitem{milyani_deep_2025}
\BIBentryALTinterwordspacing
A.~H. Milyani and E.~T. Attar, ``Deep learning for inner speech recognition: a pilot comparative study of {EEGNet} and a spectro-temporal transformer on bimodal {EEG}-{fMRI} data,'' vol.~19. [Online]. Available: \url{https://www.frontiersin.org/journals/human-neuroscience/articles/10.3389/fnhum.2025.1668935/full}
\BIBentrySTDinterwordspacing

\bibitem{umair_privacypreserving_2025}
\BIBentryALTinterwordspacing
M.~Umair, M.~S. Khan, M.~Hanif, W.~Ghaban, I.~Nafea, S.~N. Qasem, and F.~Saeed, ``Privacy–preserving dementia classification from {EEG} via hybrid–fusion {EEGNetv}4 and federated learning,'' vol.~19. [Online]. Available: \url{https://www.frontiersin.org/journals/computational-neuroscience/articles/10.3389/fncom.2025.1617883/full}
\BIBentrySTDinterwordspacing

\bibitem{otarbay_svm-enhanced_2025}
\BIBentryALTinterwordspacing
Z.~Otarbay and A.~Kyzyrkanov, ``{SVM}-enhanced attention mechanisms for motor imagery {EEG} classification in brain-computer interfaces,'' vol.~19. [Online]. Available: \url{https://www.frontiersin.org/journals/neuroscience/articles/10.3389/fnins.2025.1622847/full}
\BIBentrySTDinterwordspacing

\bibitem{ergun_attention-enhanced_2025}
\BIBentryALTinterwordspacing
E.~Ergün, ``Attention-enhanced hybrid deep learning model for robust mango leaf disease classification via {ConvNeXt} and vision transformer fusion,'' vol.~16. [Online]. Available: \url{https://www.frontiersin.org/journals/plant-science/articles/10.3389/fpls.2025.1638520/full}
\BIBentrySTDinterwordspacing

\bibitem{salman_plant_2025}
\BIBentryALTinterwordspacing
Z.~Salman, A.~Muhammad, and D.~Han, ``Plant disease classification in the wild using vision transformers and mixture of experts,'' vol.~16. [Online]. Available: \url{https://www.frontiersin.org/journals/plant-science/articles/10.3389/fpls.2025.1522985/full}
\BIBentrySTDinterwordspacing

\bibitem{tan_p4cn-yolov5s_2025}
\BIBentryALTinterwordspacing
Z.~Tan, D.~Ye, J.~Wang, and W.~Wang, ``P4cn-{YOLOv}5s: a passion fruit pests detection method based on lightweight-improved {YOLOv}5s,'' vol.~16. [Online]. Available: \url{https://www.frontiersin.org/journals/plant-science/articles/10.3389/fpls.2025.1612642/full}
\BIBentrySTDinterwordspacing

\bibitem{jha_deep_2025}
\BIBentryALTinterwordspacing
S.~K. Jha, V.~Gupta, P.~J. Sharma, A.~Mishra, and S.~Joshi, ``Deep learning super-resolution for temperature data downscaling: a comprehensive study using residual networks,'' vol.~7. [Online]. Available: \url{https://www.frontiersin.org/journals/climate/articles/10.3389/fclim.2025.1572428/full}
\BIBentrySTDinterwordspacing

\bibitem{kosmala_integrating_2018}
\BIBentryALTinterwordspacing
M.~Kosmala, K.~Hufkens, and A.~D. Richardson, ``Integrating camera imagery, crowdsourcing, and deep learning to improve high-frequency automated monitoring of snow at continental-to-global scales,'' vol.~13, no.~12, p. e0209649. [Online]. Available: \url{https://journals.plos.org/plosone/article?id=10.1371/journal.pone.0209649}
\BIBentrySTDinterwordspacing

\bibitem{miao_using_2018}
\BIBentryALTinterwordspacing
X.~Miao, H.~Miao, Y.~Jia, and Y.~Guo, ``Using a stacked-autoencoder neural network model to estimate sea state bias for a radar altimeter,'' vol.~13, no.~12, p. e0208989. [Online]. Available: \url{https://journals.plos.org/plosone/article?id=10.1371/journal.pone.0208989}
\BIBentrySTDinterwordspacing

\bibitem{lee_identification_2024}
\BIBentryALTinterwordspacing
H.~Lee, K.~Shin, Y.~Lee, S.~Lee, S.~Lee, E.~Lee, S.~W. Kim, H.~Y. Shin, J.~H. Kim, J.~Chung, and S.~Kwon, ``Identification of b cell subsets based on antigen receptor sequences using deep learning,'' vol.~15. [Online]. Available: \url{https://www.frontiersin.org/journals/immunology/articles/10.3389/fimmu.2024.1342285/full}
\BIBentrySTDinterwordspacing

\bibitem{sethna_population_2020}
\BIBentryALTinterwordspacing
Z.~Sethna, G.~Isacchini, T.~Dupic, T.~Mora, A.~M. Walczak, and Y.~Elhanati, ``Population variability in the generation and selection of t-cell repertoires,'' vol.~16, no.~12, p. e1008394. [Online]. Available: \url{https://journals.plos.org/ploscompbiol/article?id=10.1371/journal.pcbi.1008394}
\BIBentrySTDinterwordspacing

\bibitem{barreto_novel_2024}
\BIBentryALTinterwordspacing
C.~Barreto, V.~Cardoso-Jaime, and G.~Dimopoulos, ``A novel broad-spectrum antibacterial and anti-malarial anopheles gambiae cecropin promotes microbial clearance during pupation,'' vol.~20, no.~10, p. e1012652. [Online]. Available: \url{https://journals.plos.org/plospathogens/article?id=10.1371/journal.ppat.1012652}
\BIBentrySTDinterwordspacing

\bibitem{peng_multi-scale_2015}
\BIBentryALTinterwordspacing
H.-K. Peng and R.~Marculescu, ``Multi-scale compositionality: Identifying the compositional structures of social dynamics using deep learning,'' vol.~10, no.~4, p. e0118309. [Online]. Available: \url{https://journals.plos.org/plosone/article?id=10.1371/journal.pone.0118309}
\BIBentrySTDinterwordspacing

\bibitem{mu_ipso-lstm_2023}
\BIBentryALTinterwordspacing
G.~Mu, Z.~Liao, J.~Li, N.~Qin, and Z.~Yang, ``{IPSO}-{LSTM} hybrid model for predicting online public opinion trends in emergencies,'' vol.~18, no.~10, p. e0292677. [Online]. Available: \url{https://journals.plos.org/plosone/article?id=10.1371/journal.pone.0292677}
\BIBentrySTDinterwordspacing

\bibitem{yang_diffusion_2023}
\BIBentryALTinterwordspacing
F.~Yang, J.~Liu, R.~Zhang, and Y.~Yao, ``Diffusion characteristics classification framework for identification of diffusion source in complex networks,'' vol.~18, no.~5, p. e0285563. [Online]. Available: \url{https://journals.plos.org/plosone/article?id=10.1371/journal.pone.0285563}
\BIBentrySTDinterwordspacing

\bibitem{tariq_specollate_2021}
\BIBentryALTinterwordspacing
M.~U. Tariq and F.~Saeed, ``{SpeCollate}: Deep cross-modal similarity network for mass spectrometry data based peptide deductions,'' vol.~16, no.~10, p. e0259349. [Online]. Available: \url{https://journals.plos.org/plosone/article?id=10.1371/journal.pone.0259349}
\BIBentrySTDinterwordspacing

\bibitem{gueto-tettay_multienzyme_2023}
\BIBentryALTinterwordspacing
C.~Gueto-Tettay, D.~Tang, L.~Happonen, M.~Heusel, H.~Khakzad, J.~Malmström, and L.~Malmström, ``Multienzyme deep learning models improve peptide de novo sequencing by mass spectrometry proteomics,'' vol.~19, no.~1, p. e1010457. [Online]. Available: \url{https://journals.plos.org/ploscompbiol/article?id=10.1371/journal.pcbi.1010457}
\BIBentrySTDinterwordspacing

\bibitem{zhang_multimodal_2024}
\BIBentryALTinterwordspacing
W.~Zhang, N.~H. Patterson, N.~Verbeeck, J.~L. Moore, A.~Ly, R.~M. Caprioli, B.~D. Moor, J.~L. Norris, and M.~Claesen, ``Multimodal {MALDI} imaging mass spectrometry for improved diagnosis of melanoma,'' vol.~19, no.~5, p. e0304709. [Online]. Available: \url{https://journals.plos.org/plosone/article?id=10.1371/journal.pone.0304709}
\BIBentrySTDinterwordspacing

\bibitem{10.1371/journal.pone.0006393}
\BIBentryALTinterwordspacing
T.~Barnickel, J.~Weston, R.~Collobert, H.-W. Mewes, and V.~Stümpflen, ``Large scale application of neural network based semantic role labeling for automated relation extraction from biomedical texts,'' \emph{PLOS ONE}, vol.~4, no.~7, pp. 1--6, 07 2009. [Online]. Available: \url{https://doi.org/10.1371/journal.pone.0006393}
\BIBentrySTDinterwordspacing

\bibitem{qi_unified_2012}
\BIBentryALTinterwordspacing
Y.~Qi, M.~Oja, J.~Weston, and W.~S. Noble, ``A unified multitask architecture for predicting local protein properties,'' vol.~7, no.~3, p. e32235. [Online]. Available: \url{https://journals.plos.org/plosone/article?id=10.1371/journal.pone.0032235}
\BIBentrySTDinterwordspacing

\bibitem{ptms2}
\BIBentryALTinterwordspacing
J.~Kubilius, S.~Bracci, and H.~P. Op~de Beeck, ``Deep neural networks as a computational model for human shape sensitivity,'' \emph{PLOS Computational Biology}, vol.~12, no.~4, pp. 1--26, 04 2016. [Online]. Available: \url{https://doi.org/10.1371/journal.pcbi.1004896}
\BIBentrySTDinterwordspacing

\bibitem{chen_lailaps-qsm_2018}
\BIBentryALTinterwordspacing
J.~Chen, U.~Scholz, R.~Zhou, and M.~Lange, ``{LAILAPS}-{QSM}: A {RESTful} {API} and {JAVA} library for semantic query suggestions,'' vol.~14, no.~3, p. e1006058. [Online]. Available: \url{https://dx.plos.org/10.1371/journal.pcbi.1006058}
\BIBentrySTDinterwordspacing

\bibitem{bach_pixel-wise_2015}
\BIBentryALTinterwordspacing
S.~Bach, A.~Binder, G.~Montavon, F.~Klauschen, K.-R. Müller, and W.~Samek, ``On pixel-wise explanations for non-linear classifier decisions by layer-wise relevance propagation,'' vol.~10, no.~7, p. e0130140. [Online]. Available: \url{https://dx.plos.org/10.1371/journal.pone.0130140}
\BIBentrySTDinterwordspacing

\bibitem{hepburn_scientific_2026}
\BIBentryALTinterwordspacing
B.~Hepburn and H.~Andersen, ``Scientific method,'' in \emph{The Stanford Encyclopedia of Philosophy}, spring 2026~ed., E.~N. Zalta and U.~Nodelman, Eds.\hskip 1em plus 0.5em minus 0.4em\relax Metaphysics Research Lab, Stanford University. [Online]. Available: \url{https://plato.stanford.edu/archives/spr2026/entries/scientific-method/}
\BIBentrySTDinterwordspacing

\bibitem{weinberg_methods_1995}
\BIBentryALTinterwordspacing
S.~Weinberg, ``The methods of science … and those by which we live,'' vol.~8, no.~2, pp. 7--13. [Online]. Available: \url{http://link.springer.com/10.1007/BF02683184}
\BIBentrySTDinterwordspacing

\bibitem{hempel_philosophy_1966}
C.~Hempel, \emph{Philosophy of Natural Science}.\hskip 1em plus 0.5em minus 0.4em\relax Prentice Hall.

\bibitem{10.1371/journal.pone.0141287}
\BIBentryALTinterwordspacing
E.~Asgari and M.~R.~K. Mofrad, ``Continuous distributed representation of biological sequences for deep proteomics and genomics,'' \emph{PLOS ONE}, vol.~10, no.~11, pp. 1--15, 11 2015. [Online]. Available: \url{https://doi.org/10.1371/journal.pone.0141287}
\BIBentrySTDinterwordspacing

\bibitem{10.1371/journal.pcbi.1004962}
\BIBentryALTinterwordspacing
J.~Bendl, M.~Musil, J.~Štourač, J.~Zendulka, J.~Damborský, and J.~Brezovský, ``Predictsnp2: A unified platform for accurately evaluating snp effects by exploiting the different characteristics of variants in distinct genomic regions,'' \emph{PLOS Computational Biology}, vol.~12, no.~5, pp. 1--18, 05 2016. [Online]. Available: \url{https://doi.org/10.1371/journal.pcbi.1004962}
\BIBentrySTDinterwordspacing

\bibitem{10.3389/fninf.2018.00023}
\BIBentryALTinterwordspacing
D.~Wen, Z.~Wei, Y.~Zhou, G.~Li, X.~Zhang, and W.~Han, ``Deep learning methods to process fmri data and their application in the diagnosis of cognitive impairment: A brief overview and our opinion,'' \emph{Frontiers in Neuroinformatics}, vol. Volume 12 - 2018, 2018. [Online]. Available: \url{https://www.frontiersin.org/journals/neuroinformatics/articles/10.3389/fninf.2018.00023}
\BIBentrySTDinterwordspacing

\bibitem{10.1371/journal.pcbi.1010221}
\BIBentryALTinterwordspacing
C.~Capone, P.~Muratore, and P.~S. Paolucci, ``Error-based or target-based? a unified framework for learning in recurrent spiking networks,'' \emph{PLOS Computational Biology}, vol.~18, no.~6, pp. 1--18, 06 2022. [Online]. Available: \url{https://doi.org/10.1371/journal.pcbi.1010221}
\BIBentrySTDinterwordspacing

\bibitem{statisticsncsesPublicationsOutputUS2023}
\BIBentryALTinterwordspacing
{National Center for Science and Engineering}, ``Publications output: U.s. trends and international comparisons,'' number: {NSB}-2023-33. [Online]. Available: \url{https://ncses.nsf.gov/pubs/nsb202333/publication-output-by-region-country-or-economy-and-by-scientific-field}
\BIBentrySTDinterwordspacing

\bibitem{crook-semi}
\BIBentryALTinterwordspacing
O.~M. Crook, A.~Geladaki, D.~J.~H. Nightingale, O.~L. Vennard, K.~S. Lilley, L.~Gatto, and P.~D.~W. Kirk, ``A semi-supervised bayesian approach for simultaneous protein sub-cellular localisation assignment and novelty detection,'' \emph{PLOS Computational Biology}, vol.~16, no.~11, pp. 1--21, 11 2020. [Online]. Available: \url{https://doi.org/10.1371/journal.pcbi.1008288}
\BIBentrySTDinterwordspacing

\bibitem{10.3389/fspas.2023.1275516}
\BIBentryALTinterwordspacing
D.~Liu and W.~Cheng, ``Progress and prospects for research on martian topographic features and typical landform identification,'' vol. Volume 10 - 2023. [Online]. Available: \url{https://www.frontiersin.org/journals/astronomy-and-space-sciences/articles/10.3389/fspas.2023.1275516}
\BIBentrySTDinterwordspacing

\bibitem{jokinenPredictingRecognitionCell2021}
\BIBentryALTinterwordspacing
E.~Jokinen, J.~Huuhtanen, S.~Mustjoki, M.~Heinonen, and H.~Lähdesmäki, ``Predicting recognition between t cell receptors and epitopes with {TCRGP},'' vol.~17, no.~3, p. e1008814. [Online]. Available: \url{https://journals.plos.org/ploscompbiol/article?id=10.1371/journal.pcbi.1008814}
\BIBentrySTDinterwordspacing

\bibitem{zwierzynaClassificationAnalysisLarge2017}
\BIBentryALTinterwordspacing
M.~Zwierzyna and J.~P. Overington, ``Classification and analysis of a large collection of in vivo bioassay descriptions,'' vol.~13, no.~7, p. e1005641. [Online]. Available: \url{https://journals.plos.org/ploscompbiol/article?id=10.1371/journal.pcbi.1005641}
\BIBentrySTDinterwordspacing

\bibitem{rathiSmartAgricultureResource2025}
\BIBentryALTinterwordspacing
M.~Rathi and C.~Gomathy, ``Smart agriculture resource allocation and energy optimization using bidirectional long short-term memory with ant colony optimization (bi-{LSTM}–{ACO}),'' vol.~6. [Online]. Available: \url{https://www.frontiersin.org/journals/communications-and-networks/articles/10.3389/frcmn.2025.1587402/full}
\BIBentrySTDinterwordspacing

\bibitem{tsumuraInfluenceAgentsSelfdisclosure2023}
\BIBentryALTinterwordspacing
T.~Tsumura and S.~Yamada, ``Influence of agent’s self-disclosure on human empathy,'' vol.~18, no.~5, p. e0283955. [Online]. Available: \url{https://journals.plos.org/plosone/article?id=10.1371/journal.pone.0283955}
\BIBentrySTDinterwordspacing

\bibitem{dibajiSexDifferencesBrain2024}
\BIBentryALTinterwordspacing
M.~Dibaji, J.~Ospel, R.~Souza, and M.~Bento, ``Sex differences in brain {MRI} using deep learning toward fairer healthcare outcomes,'' vol.~18. [Online]. Available: \url{https://www.frontiersin.org/journals/computational-neuroscience/articles/10.3389/fncom.2024.1452457/full}
\BIBentrySTDinterwordspacing

\bibitem{dierkingDeepLearningTechniques2023}
\BIBentryALTinterwordspacing
I.~Dierking, J.~Dominguez, J.~Harbon, and J.~Heaton, ``Deep learning techniques for the localization and classification of liquid crystal phase transitions,'' vol.~3. [Online]. Available: \url{https://www.frontiersin.org/journals/soft-matter/articles/10.3389/frsfm.2023.1114551/full}
\BIBentrySTDinterwordspacing

\bibitem{miaoFeasibilityStudyDeep2025}
\BIBentryALTinterwordspacing
J.~Miao, Y.~Xu, K.~Men, and J.~Dai, ``A feasibility study of deep learning prediction model for {VMAT} patient-specific {QA},'' vol.~15. [Online]. Available: \url{https://www.frontiersin.org/journals/oncology/articles/10.3389/fonc.2025.1509449/full}
\BIBentrySTDinterwordspacing

\bibitem{pramanikMonkeypoxDetectionSkin2023}
\BIBentryALTinterwordspacing
R.~Pramanik, B.~Banerjee, G.~Efimenko, D.~Kaplun, and R.~Sarkar, ``Monkeypox detection from skin lesion images using an amalgamation of {CNN} models aided with beta function-based normalization scheme,'' vol.~18, no.~4, p. e0281815. [Online]. Available: \url{https://journals.plos.org/plosone/article?id=10.1371/journal.pone.0281815}
\BIBentrySTDinterwordspacing

\end{thebibliography}


\clearpage 
\onecolumn
\appendix 

\section*{Appendix: Examples of Filtered Natural Science Papers}
{\footnotesize
\setlength{\tabcolsep}{4pt}
\renewcommand{\arraystretch}{1.1}

\begin{longtable}{p{5cm} p{3cm} p{3cm} p{3cm} p{1.2cm}}
\caption{Example papers from our initial search query with at least one citation filtered on the natural science topics they were assigned by OpenAlex.}
\label{tab:appendix-ns-filter} \\

\toprule
\textbf{Paper Title} &
\textbf{OpenAlex Topic 1} &
\textbf{OpenAlex Topic 2} &
\textbf{OpenAlex Topic 3} &
\textbf{Filtered} \\
\midrule
\endfirsthead

\toprule
\textbf{Paper Title} &
\textbf{OpenAlex Topic 1} &
\textbf{OpenAlex Topic 2} &
\textbf{OpenAlex Topic 3} &
\textbf{Filtered} \\
\midrule
\endhead

\midrule
\multicolumn{5}{r}{Continued on next page} \\
\midrule
\endfoot

\bottomrule
\endlastfoot

Characterization of RNA polymerase II trigger loop mutations using molecular dynamics simulations and machine learning~\citep{dutagaci_characterization_2023} &
Biochemistry, Genetics and Molecular Biology &
Biochemistry, Genetics and Molecular Biology &
Biochemistry, Genetics and Molecular Biology &
True \\
\midrule

The genome of the migratory nematode, \textit{Radopholus similis}, reveals signatures of close association to the sedentary cyst nematodes~\citep{mathew_genome_2019} &
Agricultural and Biological Sciences &
Agricultural and Biological Sciences &
Agricultural and Biological Sciences &
True \\
\midrule

Generalizing biological surround suppression based on center surround similarity via deep neural network models~\citep{pan_generalizing_2023} &
Neuroscience &
Neuroscience &
Neuroscience &
True \\
\midrule

Integration of smart sensors and phytoremediation for real-time pollution monitoring and ecological restoration in agricultural waste management~\citep{guo_integration_2025} &
Environmental Science &
Environmental Science &
Environmental Science &
True \\
\midrule

A semi-supervised Bayesian approach for simultaneous protein sub-cellular localisation assignment and novelty detection~\citep{crook-semi} &
Chemistry &
Chemistry &
Biochemistry, Genetics and Molecular Biology &
True \\
\midrule

Progress and prospects for research on Martian topographic features and typical landform identification~\citep{10.3389/fspas.2023.1275516} &
Physics and Astronomy &
Physics and Astronomy &
Physics and Astronomy &
True \\
\midrule

Predicting recognition between T cell receptors and epitopes with TCRGP~\citep{jokinenPredictingRecognitionCell2021} &
Immunology and Microbiology &
Immunology and Microbiology &
Immunology and Microbiology &
True \\
\midrule

Differential transcript usage in the Parkinson’s disease brain~\citep{dick_differential_2020} &
Medicine &
Biochemistry, Genetics and Molecular Biology &
Biochemistry, Genetics and Molecular Biology &
True \\
\midrule

Classification of large ornithopod dinosaur footprints using Xception transfer learning~\citep{ha_classification_2023} &
Earth and Planetary Sciences &
Environmental Science &
Environmental Science &
True \\
\midrule

Sperm and northern bottlenose whale interactions with deep-water trawlers in the western North Atlantic~\citep{oyarbide_sperm_2023} &
Environmental Science &
Environmental Science &
Earth and Planetary Sciences &
True \\
\midrule

Machine learning identification of \textit{Pseudomonas aeruginosa} strains from colony image data~\citep{rattray_machine_2023} &
Biochemistry, Genetics and Molecular Biology &
Engineering &
Biochemistry, Genetics and Molecular Biology &
True \\
\midrule

A novel broad-spectrum antibacterial and anti-malarial \textit{Anopheles gambiae} Cecropin promotes microbial clearance during pupation~\citep{barreto_novel_2024} &
Immunology and Microbiology &
Immunology and Microbiology &
Agricultural and Biological Sciences &
True \\
\midrule

Multienzyme deep learning models improve peptide de novo sequencing by mass spectrometry proteomics~\citep{gueto-tettay_multienzyme_2023} &
Chemistry &
Chemistry &
Biochemistry, Genetics and Molecular Biology &
True \\
\midrule

Using a stacked-autoencoder neural network model to estimate sea state bias for a radar altimeter~\citep{miao_using_2018} &
Earth and Planetary Sciences &
Earth and Planetary Sciences &
Environmental Science &
True \\
\midrule

Machine learning methods to predict particulate matter PM$_{2.5}$~\citep{palanichamy_machine_2022} &
Environmental Science &
Environmental Science &
Engineering &
True \\
\midrule

Fish keypoint detection for offshore aquaculture: a robust deep learning approach with PCA-based shape constraint~\citep{li_fish_2025} &
Environmental Science &
Environmental Science &
Biochemistry, Genetics and Molecular Biology &
True \\
\midrule

Population variability in the generation and selection of T-cell repertoires~\citep{sethna_population_2020} &
Immunology and Microbiology &
Immunology and Microbiology &
Medicine &
True \\
\midrule

Identification of B cell subsets based on antigen receptor sequences using deep learning~\citep{lee_identification_2024} &
Immunology and Microbiology &
Immunology and Microbiology &
Biochemistry, Genetics and Molecular Biology &
True \\
\midrule

SVM-enhanced attention mechanisms for motor imagery EEG classification in brain-computer interfaces~\citep{otarbay_svm-enhanced_2025} &
Neuroscience &
Neuroscience &
Engineering &
True \\
\midrule

Multi-Scale Compositionality: Identifying the Compositional Structures of Social Dynamics Using Deep Learning~\citep{peng_multi-scale_2015} &
Physics and Astronomy &
Physics and Astronomy &
Computer Science &
True \\
\midrule

P4CN-YOLOv5s: a passion fruit pests detection method based on lightweight-improved YOLOv5s &
Agricultural and Biological Sciences~\citep{tan_p4cn-yolov5s_2025} &
Agricultural and Biological Sciences &
Biochemistry, Genetics and Molecular Biology &
True \\
\midrule

Classification and analysis of a large collection of in vivo bioassay descriptions~\citep{zwierzynaClassificationAnalysisLarge2017} &
Biochemistry, Genetics and Molecular Biology &
Computer Science &
Veterinary &
False \\
\midrule

Smart agriculture resource allocation and energy optimization using bidirectional long short-term memory with ant colony optimization (Bi-LSTM--ACO)~\citep{rathiSmartAgricultureResource2025} &
Agricultural and Biological Sciences &
Engineering &
Computer Science &
False \\
\midrule

Influence of agent’s self-disclosure on human empathy~\citep{tsumuraInfluenceAgentsSelfdisclosure2023} &
Neuroscience &
Psychology &
Social Sciences &
False \\
\midrule

Sex differences in brain MRI using deep learning toward fairer healthcare outcomes~\citep{dibajiSexDifferencesBrain2024} &
Environmental Science &
Medicine &
Social Sciences &
False \\
\midrule

Deep learning techniques for the localization and classification of liquid crystal phase transitions~\citep{dierkingDeepLearningTechniques2023} &
Chemistry &
Materials Science &
Economics, Econometrics and Finance &
False \\
\midrule

A feasibility study of deep learning prediction model for VMAT patient-specific QA~\citep{miaoFeasibilityStudyDeep2025} &
Physics and Astronomy &
Medicine &
Engineering &
False \\
\midrule

Monkeypox detection from skin lesion images using an amalgamation of CNN models aided with Beta function-based normalization scheme~\citep{pramanikMonkeypoxDetectionSkin2023} &
Immunology and Microbiology &
Computer Science &
Computer Science &
False \\

\end{longtable}
}
\newpage

\section*{Appendix: GPT-OSS Prompts} 

\definecolor{myboxbg}{RGB}{248,248,248}
\definecolor{myboxframe}{RGB}{90,90,90}

\tcbset{
  mypromptbox/.style={
    enhanced,
    breakable,
    colback=myboxbg,
    colframe=myboxframe,
    boxrule=0.6pt,
    arc=2mm,
    left=3mm,
    right=3mm,
    top=2mm,
    bottom=2mm,
    fonttitle=\bfseries,
    title=#1,
    before skip=8pt,
    after skip=8pt
  }
}

\begin{tcolorbox}[mypromptbox={Prompt 1: Identify Deep Learning Usage In Natural Science Works}]

\textbf{Context}
You are an AI model integrated into an automated pipeline that processes academic computational Natural Science papers into a machine-readable format. Your sole responsibility is to evaluate the paper’s content and determine whether the author uses deep learning models or methods in their methodology. Your response will be consumed by downstream systems that require structured JSON.

\textbf{Objective}
Your task is to output only a JSON object containing key-value pairs, where:
\begin{itemize}
    \item The key ``result'' value is a boolean (\texttt{true} or \texttt{false}) based on whether the input text uses deep learning models or methods in its methodology, and
    \item The key ``prose'' value is the most salient excerpt from the paper that shows concrete evidence of deep learning usage in the paper, or empty if no deep learning methods are used.
\end{itemize}
No explanations or extra output are allowed.

\textbf{Style}
Responses must be strictly machine-readable JSON. No natural language,
commentary, or formatting beyond the JSON object is permitted.

\textbf{Tone}
Neutral, objective, and machine-like.

\textbf{Audience}
The audience is a machine system that parses JSON. Human readability is
irrelevant.

\textbf{Response}
Return only a JSON object of the form:

\begin{verbatim}
{
    "result": boolean,
    "prose": string | null
}
\end{verbatim}

Nothing else should ever be returned.
\end{tcolorbox}

\definecolor{myboxbg}{RGB}{248,248,248}
\definecolor{myboxframe}{RGB}{90,90,90}

\tcbset{
  mypromptbox/.style={
    enhanced,
    breakable,
    colback=myboxbg,
    colframe=myboxframe,
    boxrule=0.6pt,
    arc=2mm,
    left=3mm,
    right=3mm,
    top=2mm,
    bottom=2mm,
    fonttitle=\bfseries,
    title=#1,
    before skip=8pt,
    after skip=8pt
  }
}

\begin{tcolorbox}[mypromptbox={Prompt 2: dentify Pre-Trained Model Usage In Deep Learning Using Natural Science Works}]

\textbf{Context}
You are an AI model integrated into an automated pipeline that processes academic Computational Natural Science papers into a machine-readable format. Your sole responsibility is to evaluate the paper's content and determine whether the authors use pre-trained deep learning models (PTMs) in their methodology. Your response will be consumed by downstream systems that require a structured JSON format.

\textbf{Objective}
Your task is to output only a JSON object containing key-value pairs, where:
\begin{itemize}
    \item The key ``result'' value is a boolean (\texttt{true} or \texttt{false}) based on whether the input text indicates the use of pre-trained deep learning models (PTMs) in the methodology, and
    \item The key ``prose'' value is the most salient excerpt from the paper that shows concrete evidence of pre-trained model usage, or an empty string if no PTMs are used.
\end{itemize}
No explanations or extra output are allowed.

\textbf{Style}
Responses must be strictly machine-readable JSON. No natural language,
commentary, or formatting beyond the JSON object is permitted.

\textbf{Tone}
Neutral, objective, and machine-like.

\textbf{Audience}
The audience is a machine system that parses JSON. Human readability is
irrelevant.

\textbf{Response}
Return only a JSON object of the form:

\begin{verbatim}
{
    "result": boolean,
    "prose": string | null
}
\end{verbatim}

Nothing else should ever be returned.
\end{tcolorbox}
\clearpage

\definecolor{myboxbg}{RGB}{248,248,248}
\definecolor{myboxframe}{RGB}{90,90,90}

\tcbset{
  mypromptbox/.style={
    enhanced,
    breakable,
    colback=myboxbg,
    colframe=myboxframe,
    boxrule=0.6pt,
    arc=2mm,
    left=3mm,
    right=3mm,
    top=2mm,
    bottom=2mm,
    fonttitle=\bfseries,
    title=#1,
    before skip=8pt,
    after skip=8pt
  }
}

\begin{tcolorbox}[mypromptbox={Prompt 3: Identify Specific Pre-Trained Models In Pre-Trained Model Using Natural Science Works}]

\textbf{Context}
You are an AI model integrated into an automated pipeline that processes academic Computational Natural Science papers into a machine-readable format. Your sole responsibility is to evaluate the paper's content and determine what pre-trained deep learning models the authors use in their methodology. Your response will be consumed by downstream systems that require a structured JSON format.

Pre-trained deep learning models have many different names. The following is a list of pre-trained deep learning model names and their data modality that you can reference in your analysis:

\begin{verbatim}
{
  "text_models": [
    "ALBERT", "Apertus", "Arcee", "Bamba", "BART", "BARThez", "BARTpho", "BERT", "BertGeneration",
    "BertJapanese", "BERTweet", "BigBird", "BigBirdPegasus", "BioGpt", "BitNet", "Blenderbot",
    "Blenderbot Small", "BLOOM", "BLT", "BORT", "ByT5", "CamemBERT", "CANINE", "CodeGen",
    "CodeLlama", "Cohere", "Cohere2", "ConvBERT", "CPM", "CPMANT", "CTRL", "DBRX", "DeBERTa",
    "DeBERTa-v2", "DeepSeek-V2", "DeepSeek-V3", "DialoGPT", "DiffLlama", "DistilBERT", "Doge",
    "dots1", "DPR", "ELECTRA", "Encoder Decoder Models", "ERNIE", "Ernie4_5", "Ernie4_5_MoE",
    "ErnieM", "ESM", "EXAONE-4.0", "Falcon", "Falcon3", "FalconH1", "FalconMamba", "FLAN-T5",
    "FLAN-UL2", "FlauBERT", "FlexOlmo", "FNet", "FSMT", "Funnel Transformer", "Fuyu", "Gemma",
    "Gemma2", "GLM", "glm4", "glm4_moe", "GPT", "GPT Neo", "GPT NeoX", "GPT NeoX Japanese",
    "GPT-J", "GPT2", "GPTBigCode", "GptOss", "GPTSAN Japanese", "GPTSw3", "Granite", "GraniteMoe",
    "GraniteMoeHybrid", "GraniteMoeShared", "Helium", "HerBERT", "HGNet-V2", "HunYuanDenseV1",
    "HunYuanMoEV1", "I-BERT", "Jamba", "JetMoe", "Jukebox", "LED", "LFM2", "LLaMA", "Llama2",
    "Llama3", "LongCatFlash", "Longformer", "LongT5", "LUKE", "M2M100", "MADLAD-400", "Mamba",
    "Mamba2", "MarianMT", "MarkupLM", "MBart and MBart-50", "MEGA", "MegatronBERT",
    "MegatronGPT2", "MiniMax", "Ministral", "Mistral", "Mixtral", "mLUKE", "MobileBERT",
    "ModernBert", "ModernBERTDecoder", "MPNet", "MPT", "MRA", "MT5", "MVP", "myt5", "Nemotron",
    "NEZHA", "NLLB", "NLLB-MoE", "Nyströmformer", "OLMo", "OLMo2", "Olmo3", "OLMoE",
    "Open-Llama", "OPT", "Pegasus", "PEGASUS-X", "Persimmon", "Phi", "Phi-3", "PhiMoE",
    "PhoBERT", "PLBart", "ProphetNet", "QDQBert", "Qwen2", "Qwen2MoE", "Qwen3", "Qwen3MoE",
    "Qwen3Next", "RAG", "REALM", "RecurrentGemma", "Reformer", "RemBERT", "RetriBERT",
    "RoBERTa", "RoBERTa-PreLayerNorm", "RoCBert", "RoFormer", "RWKV", "Seed-Oss", "Splinter",
    "SqueezeBERT", "StableLm", "Starcoder2", "SwitchTransformers", "T5", "T5Gemma",
    "T5v1.1", "TAPEX", "Transformer XL", "UL2", "UMT5", "VaultGemma", "X-MOD", "XGLM",
    "XLM", "XLM-ProphetNet", "XLM-RoBERTa", "XLM-RoBERTa-XL", "XLM-V", "XLNet", "xLSTM",
    "YOSO", "Zamba", "Zamba2"
  ],
  "vision_models": [
    "Aimv2", "BEiT", "BiT", "Conditional DETR", "ConvNeXT", "ConvNeXTV2", "CvT", "D-FINE",
    "DAB-DETR", "Deformable DETR", "DeiT", "Depth Anything", "Depth Anything V2",
    "DepthPro", "DETA", "DETR", "DiNAT", "DINOV2", "DINOv2 with Registers", "DINOv3",
    "DiT", "DPT", "EfficientFormer", "EfficientLoFTR", "EfficientNet", "EoMT",
    "FocalNet", "GLPN", "HGNet-V2", "Hiera", "I-JEPA", "ImageGPT", "LeViT", "LightGlue",
    "Mask2Former", "MaskFormer", "MLCD", "MobileNetV1", "MobileNetV2", "MobileViT",
    "MobileViTV2", "NAT", "PoolFormer", "Prompt Depth Anything",
    "Pyramid Vision Transformer (PVT)",
    "Pyramid Vision Transformer v2 (PVTv2)", "RegNet", "ResNet", "RT-DETR",
    "RT-DETRv2", "SAM2", "SegFormer", "SegGpt", "Segment Anything",
    "Segment Anything High Quality", "SuperGlue", "SuperPoint", "SwiftFormer",
    "Swin Transformer", "Swin Transformer V2", "Swin2SR", "Table Transformer",
    "TextNet", "Timm Wrapper", "UperNet", "VAN", "Vision Transformer (ViT)",
    "ViT Hybrid", "ViTDet", "ViTMAE", "ViTMatte", "ViTMSN", "ViTPose",
    "YOLOS", "ZoeDepth"
  ],
  "audio_models": [
    "Audio Spectrogram Transformer", "Bark", "CLAP", "CSM", "dac", "Dia", "EnCodec",
    "FastSpeech2Conformer", "GraniteSpeech", "Hubert", "Kyutai Speech-To-Text",
    "MCTCT", "Mimi", "MMS", "Moonshine", "Moshi", "MusicGen", "MusicGen Melody",
    "Parakeet", "Pop2Piano", "Seamless-M4T", "SeamlessM4T-v2", "SEW", "SEW-D",
    "Speech2Text", "Speech2Text2", "SpeechT5", "UniSpeech", "UniSpeech-SAT",
    "UnivNet", "VITS", "Wav2Vec2", "Wav2Vec2-BERT", "Wav2Vec2-Conformer",
    "Wav2Vec2Phoneme", "WavLM", "Whisper", "X-Codec", "XLS-R", "XLSR-Wav2Vec2"
  ],
  "video_models": [
    "SAM2 Video", "TimeSformer", "V-JEPA 2", "VideoMAE", "ViViT"
  ],
  "multimodal_models": [
    "ALIGN", "AltCLIP", "Aria", "AyaVision", "BLIP", "BLIP-2", "BridgeTower", "BROS",
    "Chameleon", "Chinese-CLIP", "CLIP", "CLIPSeg", "CLVP", "Cohere2Vision",
    "ColPali", "ColQwen2", "Data2Vec", "DeepseekVL", "DeepseekVLHybrid", "DePlot",
    "Donut", "EdgeTAM", "EdgeTamVideo", "Emu3", "Evolla", "FLAVA", "Florence2",
    "Gemma3", "Gemma3n", "GIT", "glm4v", "glm4v_moe", "GOT-OCR2", "GraniteVision",
    "Grounding", "DINO", "GroupViT", "IDEFICS", "Idefics2", "Idefics3",
    "InstructBLIP", "InstructBlipVideo", "InternVL", "Janus", "KOSMOS-2",
    "KOSMOS-2.5", "LayoutLM", "LayoutLMV2", "LayoutLMV3", "LayoutXLM",
    "LFM2-VL", "LiLT", "Llama4", "LLaVA", "LLaVA-NeXT", "LLaVa-NeXT-Video",
    "LLaVA-Onevision", "LXMERT", "MatCha", "MetaCLIP 2", "MGP-STR",
    "Mistral3", "mllama", "MM Grounding DINO", "Nougat", "OmDet-Turbo",
    "OneFormer", "Ovis2", "OWL-ViT", "OWLv2", "PaliGemma", "Perceiver",
    "PerceptionLM", "Phi4", "Multimodal", "Pix2Struct", "Pixtral",
    "Qwen2.5-Omni", "Qwen2.5-VL", "Qwen2Audio", "Qwen2VL",
    "Qwen3-Omni-MoE", "Qwen3VL", "Qwen3VLMoe", "ShieldGemma2",
    "SigLIP", "SigLIP2", "SmolLM3", "SmolVLM",
    "Speech Encoder Decoder Models", "TAPAS", "TrOCR", "TVLT", "TVP",
    "UDOP", "VideoLlava", "ViLT", "VipLlava",
    "Vision Encoder Decoder Models", "Vision Text Dual Encoder",
    "VisualBERT", "Voxtral", "X-CLIP"
  ],
  "reinforcement_learning_models": [
    "Decision Transformer", "Trajectory Transformer"
  ],
  "time_series_models": [
    "Autoformer", "Informer", "PatchTSMixer", "PatchTST",
    "Time Series Transformer", "TimesFM"
  ],
  "graph_models": [
    "Graphormer"
  ],
  "edge_cases": [
    "SignalP", "DeepMedic", "SecretomeP", "DeepLoc", "DeepLocPro"
  ]
}
\end{verbatim}

However, not all papers use these models. In the event that the paper does not
use one of these PTMs, identify the PTM.

\textbf{Objective}
Your task is to output only an array of JSON objects containing key-value pairs, where:
\begin{itemize}
    \item The key ``model'' value is a string of the pre-trained deep learning model's name as stated in the prose, and
    \item The key ``prose'' value is the most salient excerpt from the paper that shows concrete evidence of the pre-trained model's usage.
\end{itemize}
No explanations or extra output are allowed.

\textbf{Style}
Responses must be strictly machine-readable JSON. No natural language,
commentary, or formatting beyond the JSON object is permitted.

\textbf{Tone}
Neutral, objective, and machine-like.

\textbf{Audience}
The audience is a machine system that parses JSON. Human readability is
irrelevant.

\textbf{Response}
Return only an array of JSON objects of the form:

\begin{verbatim}
[
    {
        "model": string,
        "prose": string
    },
    ...
]
\end{verbatim}

Nothing else should ever be returned.
\end{tcolorbox}
\clearpage

\definecolor{myboxbg}{RGB}{248,248,248}
\definecolor{myboxframe}{RGB}{90,90,90}

\tcbset{
  mypromptbox/.style={
    enhanced,
    breakable,
    colback=myboxbg,
    colframe=myboxframe,
    boxrule=0.6pt,
    arc=2mm,
    left=3mm,
    right=3mm,
    top=2mm,
    bottom=2mm,
    fonttitle=\bfseries,
    title=#1,
    before skip=8pt,
    after skip=8pt
  }
}

\begin{tcolorbox}[mypromptbox={Prompt 4: Identify Pre-Trained Model Reuse Paradigm In Pre-Trained Model Using Natural Science Works}]

\textbf{Context}
You are an AI model integrated into an automated pipeline that processes academic Computational Natural Science papers into a machine-readable format. Your sole responsibility is to evaluate the paper's content and determine the form and classification of pre-trained deep learning reuse models that the authors use in their methodology. Your response will be consumed by downstream systems that require a structured JSON format.

Pre-trained deep learning models have many different names. The following is a list of pre-trained deep learning model names and their data modality that you can reference in your analysis:

\begin{verbatim}
{
  "text_models": [
    "ALBERT", "Apertus", "Arcee", "Bamba", "BART", "BARThez", "BARTpho", "BERT", "BertGeneration",
    "BertJapanese", "BERTweet", "BigBird", "BigBirdPegasus", "BioGpt", "BitNet", "Blenderbot",
    "Blenderbot Small", "BLOOM", "BLT", "BORT", "ByT5", "CamemBERT", "CANINE", "CodeGen",
    "CodeLlama", "Cohere", "Cohere2", "ConvBERT", "CPM", "CPMANT", "CTRL", "DBRX", "DeBERTa",
    "DeBERTa-v2", "DeepSeek-V2", "DeepSeek-V3", "DialoGPT", "DiffLlama", "DistilBERT", "Doge",
    "dots1", "DPR", "ELECTRA", "Encoder Decoder Models", "ERNIE", "Ernie4_5", "Ernie4_5_MoE",
    "ErnieM", "ESM", "EXAONE-4.0", "Falcon", "Falcon3", "FalconH1", "FalconMamba", "FLAN-T5",
    "FLAN-UL2", "FlauBERT", "FlexOlmo", "FNet", "FSMT", "Funnel Transformer", "Fuyu", "Gemma",
    "Gemma2", "GLM", "glm4", "glm4_moe", "GPT", "GPT Neo", "GPT NeoX", "GPT NeoX Japanese",
    "GPT-J", "GPT2", "GPTBigCode", "GptOss", "GPTSAN Japanese", "GPTSw3", "Granite", "GraniteMoe",
    "GraniteMoeHybrid", "GraniteMoeShared", "Helium", "HerBERT", "HGNet-V2", "HunYuanDenseV1",
    "HunYuanMoEV1", "I-BERT", "Jamba", "JetMoe", "Jukebox", "LED", "LFM2", "LLaMA", "Llama2",
    "Llama3", "LongCatFlash", "Longformer", "LongT5", "LUKE", "M2M100", "MADLAD-400", "Mamba",
    "Mamba2", "MarianMT", "MarkupLM", "MBart and MBart-50", "MEGA", "MegatronBERT",
    "MegatronGPT2", "MiniMax", "Ministral", "Mistral", "Mixtral", "mLUKE", "MobileBERT",
    "ModernBert", "ModernBERTDecoder", "MPNet", "MPT", "MRA", "MT5", "MVP", "myt5", "Nemotron",
    "NEZHA", "NLLB", "NLLB-MoE", "Nyströmformer", "OLMo", "OLMo2", "Olmo3", "OLMoE",
    "Open-Llama", "OPT", "Pegasus", "PEGASUS-X", "Persimmon", "Phi", "Phi-3", "PhiMoE",
    "PhoBERT", "PLBart", "ProphetNet", "QDQBert", "Qwen2", "Qwen2MoE", "Qwen3", "Qwen3MoE",
    "Qwen3Next", "RAG", "REALM", "RecurrentGemma", "Reformer", "RemBERT", "RetriBERT",
    "RoBERTa", "RoBERTa-PreLayerNorm", "RoCBert", "RoFormer", "RWKV", "Seed-Oss", "Splinter",
    "SqueezeBERT", "StableLm", "Starcoder2", "SwitchTransformers", "T5", "T5Gemma",
    "T5v1.1", "TAPEX", "Transformer XL", "UL2", "UMT5", "VaultGemma", "X-MOD", "XGLM",
    "XLM", "XLM-ProphetNet", "XLM-RoBERTa", "XLM-RoBERTa-XL", "XLM-V", "XLNet", "xLSTM",
    "YOSO", "Zamba", "Zamba2"
  ],
  "vision_models": [
    "Aimv2", "BEiT", "BiT", "Conditional DETR", "ConvNeXT", "ConvNeXTV2", "CvT", "D-FINE",
    "DAB-DETR", "Deformable DETR", "DeiT", "Depth Anything", "Depth Anything V2",
    "DepthPro", "DETA", "DETR", "DiNAT", "DINOV2", "DINOv2 with Registers", "DINOv3",
    "DiT", "DPT", "EfficientFormer", "EfficientLoFTR", "EfficientNet", "EoMT",
    "FocalNet", "GLPN", "HGNet-V2", "Hiera", "I-JEPA", "ImageGPT", "LeViT", "LightGlue",
    "Mask2Former", "MaskFormer", "MLCD", "MobileNetV1", "MobileNetV2", "MobileViT",
    "MobileViTV2", "NAT", "PoolFormer", "Prompt Depth Anything",
    "Pyramid Vision Transformer (PVT)",
    "Pyramid Vision Transformer v2 (PVTv2)", "RegNet", "ResNet", "RT-DETR",
    "RT-DETRv2", "SAM2", "SegFormer", "SegGpt", "Segment Anything",
    "Segment Anything High Quality", "SuperGlue", "SuperPoint", "SwiftFormer",
    "Swin Transformer", "Swin Transformer V2", "Swin2SR", "Table Transformer",
    "TextNet", "Timm Wrapper", "UperNet", "VAN", "Vision Transformer (ViT)",
    "ViT Hybrid", "ViTDet", "ViTMAE", "ViTMatte", "ViTMSN", "ViTPose",
    "YOLOS", "ZoeDepth"
  ],
  "audio_models": [
    "Audio Spectrogram Transformer", "Bark", "CLAP", "CSM", "dac", "Dia", "EnCodec",
    "FastSpeech2Conformer", "GraniteSpeech", "Hubert", "Kyutai Speech-To-Text",
    "MCTCT", "Mimi", "MMS", "Moonshine", "Moshi", "MusicGen", "MusicGen Melody",
    "Parakeet", "Pop2Piano", "Seamless-M4T", "SeamlessM4T-v2", "SEW", "SEW-D",
    "Speech2Text", "Speech2Text2", "SpeechT5", "UniSpeech", "UniSpeech-SAT",
    "UnivNet", "VITS", "Wav2Vec2", "Wav2Vec2-BERT", "Wav2Vec2-Conformer",
    "Wav2Vec2Phoneme", "WavLM", "Whisper", "X-Codec", "XLS-R", "XLSR-Wav2Vec2"
  ],
  "video_models": [
    "SAM2 Video", "TimeSformer", "V-JEPA 2", "VideoMAE", "ViViT"
  ],
  "multimodal_models": [
    "ALIGN", "AltCLIP", "Aria", "AyaVision", "BLIP", "BLIP-2", "BridgeTower", "BROS",
    "Chameleon", "Chinese-CLIP", "CLIP", "CLIPSeg", "CLVP", "Cohere2Vision",
    "ColPali", "ColQwen2", "Data2Vec", "DeepseekVL", "DeepseekVLHybrid", "DePlot",
    "Donut", "EdgeTAM", "EdgeTamVideo", "Emu3", "Evolla", "FLAVA", "Florence2",
    "Gemma3", "Gemma3n", "GIT", "glm4v", "glm4v_moe", "GOT-OCR2", "GraniteVision",
    "Grounding", "DINO", "GroupViT", "IDEFICS", "Idefics2", "Idefics3",
    "InstructBLIP", "InstructBlipVideo", "InternVL", "Janus", "KOSMOS-2",
    "KOSMOS-2.5", "LayoutLM", "LayoutLMV2", "LayoutLMV3", "LayoutXLM",
    "LFM2-VL", "LiLT", "Llama4", "LLaVA", "LLaVA-NeXT", "LLaVa-NeXT-Video",
    "LLaVA-Onevision", "LXMERT", "MatCha", "MetaCLIP 2", "MGP-STR",
    "Mistral3", "mllama", "MM Grounding DINO", "Nougat", "OmDet-Turbo",
    "OneFormer", "Ovis2", "OWL-ViT", "OWLv2", "PaliGemma", "Perceiver",
    "PerceptionLM", "Phi4", "Multimodal", "Pix2Struct", "Pixtral",
    "Qwen2.5-Omni", "Qwen2.5-VL", "Qwen2Audio", "Qwen2VL",
    "Qwen3-Omni-MoE", "Qwen3VL", "Qwen3VLMoe", "ShieldGemma2",
    "SigLIP", "SigLIP2", "SmolLM3", "SmolVLM",
    "Speech Encoder Decoder Models", "TAPAS", "TrOCR", "TVLT", "TVP",
    "UDOP", "VideoLlava", "ViLT", "VipLlava",
    "Vision Encoder Decoder Models", "Vision Text Dual Encoder",
    "VisualBERT", "Voxtral", "X-CLIP"
  ],
  "reinforcement_learning_models": [
    "Decision Transformer", "Trajectory Transformer"
  ],
  "time_series_models": [
    "Autoformer", "Informer", "PatchTSMixer", "PatchTST",
    "Time Series Transformer", "TimesFM"
  ],
  "graph_models": [
    "Graphormer"
  ],
  "edge_cases": [
    "SignalP", "DeepMedic", "SecretomeP", "DeepLoc", "DeepLocPro"
  ]
}
\end{verbatim}
However, not all papers use these models. In the event that the paper does not use one of these PTMs, identify the PTM.

Pre-trained deep learning reuse methods can be classified as one of the following:

\begin{itemize}
    \item \textbf{Conceptual Reuse:} Replicate and reengineer algorithms or architectures (e.g., independently confirming results or reimplementing in a different framework).
    \item \textbf{Adaptation Reuse:} Leverage existing DNN models and adapt them to solve different tasks (e.g., transfer learning, fine-tuning, or LoRA).
    \item \textbf{Deployment Reuse:} Convert and deploy pre-trained models in different environments (e.g., quantization, ONNX conversion, or inference server deployment).
\end{itemize}

\textbf{Objective}
Output only an array of JSON objects containing:
\begin{itemize}
    \item \textbf{model}: Name of the pre-trained model as stated in the prose.
    \item \textbf{form}: The specific reuse form described.
    \item \textbf{classification}: One of: \texttt{"conceptual\_reuse"}, \texttt{"adaptation\_reuse"}, or \texttt{"deployment\_reuse"}.
    \item \textbf{prose}: The salient excerpt showing concrete evidence.
\end{itemize}

\textbf{Style}
Responses must be strictly machine-readable JSON. No natural language,
commentary, or formatting beyond the JSON object is permitted.

\textbf{Tone}
Neutral, objective, and machine-like.

\textbf{Audience}
The audience is a machine system that parses JSON. Human readability is
irrelevant.

\textbf{Response}
Return only an array of JSON objects of the form:

\begin{verbatim}
[
    {
        "model": string,
        "form": string,
        "classification": string,
        "prose": string
    },
    ...
]
\end{verbatim}

Nothing else should ever be returned.
\end{tcolorbox}
\clearpage

\definecolor{myboxbg}{RGB}{248,248,248}
\definecolor{myboxframe}{RGB}{90,90,90}

\tcbset{
  mypromptbox/.style={
    enhanced,
    breakable,
    colback=myboxbg,
    colframe=myboxframe,
    boxrule=0.6pt,
    arc=2mm,
    left=3mm,
    right=3mm,
    top=2mm,
    bottom=2mm,
    fonttitle=\bfseries,
    title=#1,
    before skip=8pt,
    after skip=8pt
  }
}

\begin{tcolorbox}[mypromptbox={Prompt 5: Identify Pre-Trained Model Reuse Impact In Pre-Trained Model Using Natural Science Works}]

\textbf{Context}
You are an AI model integrated into an automated pipeline that processes academic Computational Natural Science papers into a machine-readable format. Your sole responsibility is to evaluate the paper's content and determine where in the scientific method a pre-trained deep learning model was leveraged. Your response will be consumed by downstream systems that require a structured JSON format.

\begin{verbatim}
{
  "text_models": [
    "ALBERT", "Apertus", "Arcee", "Bamba", "BART", "BARThez", "BARTpho", "BERT", "BertGeneration",
    "BertJapanese", "BERTweet", "BigBird", "BigBirdPegasus", "BioGpt", "BitNet", "Blenderbot",
    "Blenderbot Small", "BLOOM", "BLT", "BORT", "ByT5", "CamemBERT", "CANINE", "CodeGen",
    "CodeLlama", "Cohere", "Cohere2", "ConvBERT", "CPM", "CPMANT", "CTRL", "DBRX", "DeBERTa",
    "DeBERTa-v2", "DeepSeek-V2", "DeepSeek-V3", "DialoGPT", "DiffLlama", "DistilBERT", "Doge",
    "dots1", "DPR", "ELECTRA", "Encoder Decoder Models", "ERNIE", "Ernie4_5", "Ernie4_5_MoE",
    "ErnieM", "ESM", "EXAONE-4.0", "Falcon", "Falcon3", "FalconH1", "FalconMamba", "FLAN-T5",
    "FLAN-UL2", "FlauBERT", "FlexOlmo", "FNet", "FSMT", "Funnel Transformer", "Fuyu", "Gemma",
    "Gemma2", "GLM", "glm4", "glm4_moe", "GPT", "GPT Neo", "GPT NeoX", "GPT NeoX Japanese",
    "GPT-J", "GPT2", "GPTBigCode", "GptOss", "GPTSAN Japanese", "GPTSw3", "Granite", "GraniteMoe",
    "GraniteMoeHybrid", "GraniteMoeShared", "Helium", "HerBERT", "HGNet-V2", "HunYuanDenseV1",
    "HunYuanMoEV1", "I-BERT", "Jamba", "JetMoe", "Jukebox", "LED", "LFM2", "LLaMA", "Llama2",
    "Llama3", "LongCatFlash", "Longformer", "LongT5", "LUKE", "M2M100", "MADLAD-400", "Mamba",
    "Mamba2", "MarianMT", "MarkupLM", "MBart and MBart-50", "MEGA", "MegatronBERT",
    "MegatronGPT2", "MiniMax", "Ministral", "Mistral", "Mixtral", "mLUKE", "MobileBERT",
    "ModernBert", "ModernBERTDecoder", "MPNet", "MPT", "MRA", "MT5", "MVP", "myt5", "Nemotron",
    "NEZHA", "NLLB", "NLLB-MoE", "Nyströmformer", "OLMo", "OLMo2", "Olmo3", "OLMoE",
    "Open-Llama", "OPT", "Pegasus", "PEGASUS-X", "Persimmon", "Phi", "Phi-3", "PhiMoE",
    "PhoBERT", "PLBart", "ProphetNet", "QDQBert", "Qwen2", "Qwen2MoE", "Qwen3", "Qwen3MoE",
    "Qwen3Next", "RAG", "REALM", "RecurrentGemma", "Reformer", "RemBERT", "RetriBERT",
    "RoBERTa", "RoBERTa-PreLayerNorm", "RoCBert", "RoFormer", "RWKV", "Seed-Oss", "Splinter",
    "SqueezeBERT", "StableLm", "Starcoder2", "SwitchTransformers", "T5", "T5Gemma",
    "T5v1.1", "TAPEX", "Transformer XL", "UL2", "UMT5", "VaultGemma", "X-MOD", "XGLM",
    "XLM", "XLM-ProphetNet", "XLM-RoBERTa", "XLM-RoBERTa-XL", "XLM-V", "XLNet", "xLSTM",
    "YOSO", "Zamba", "Zamba2"
  ],

  "vision_models": [
    "Aimv2", "BEiT", "BiT", "Conditional DETR", "ConvNeXT", "ConvNeXTV2", "CvT", "D-FINE",
    "DAB-DETR", "Deformable DETR", "DeiT", "Depth Anything", "Depth Anything V2",
    "DepthPro", "DETA", "DETR", "DiNAT", "DINOV2", "DINOv2 with Registers", "DINOv3",
    "DiT", "DPT", "EfficientFormer", "EfficientLoFTR", "EfficientNet", "EoMT",
    "FocalNet", "GLPN", "HGNet-V2", "Hiera", "I-JEPA", "ImageGPT", "LeViT", "LightGlue",
    "Mask2Former", "MaskFormer", "MLCD", "MobileNetV1", "MobileNetV2", "MobileViT",
    "MobileViTV2", "NAT", "PoolFormer", "Prompt Depth Anything",
    "Pyramid Vision Transformer (PVT)",
    "Pyramid Vision Transformer v2 (PVTv2)", "RegNet", "ResNet", "RT-DETR",
    "RT-DETRv2", "SAM2", "SegFormer", "SegGpt", "Segment Anything",
    "Segment Anything High Quality", "SuperGlue", "SuperPoint", "SwiftFormer",
    "Swin Transformer", "Swin Transformer V2", "Swin2SR", "Table Transformer",
    "TextNet", "Timm Wrapper", "UperNet", "VAN", "Vision Transformer (ViT)",
    "ViT Hybrid", "ViTDet", "ViTMAE", "ViTMatte", "ViTMSN", "ViTPose",
    "YOLOS", "ZoeDepth"
  ],

  "audio_models": [
    "Audio Spectrogram Transformer", "Bark", "CLAP", "CSM", "dac", "Dia", "EnCodec",
    "FastSpeech2Conformer", "GraniteSpeech", "Hubert", "Kyutai Speech-To-Text",
    "MCTCT", "Mimi", "MMS", "Moonshine", "Moshi", "MusicGen", "MusicGen Melody",
    "Parakeet", "Pop2Piano", "Seamless-M4T", "SeamlessM4T-v2", "SEW", "SEW-D",
    "Speech2Text", "Speech2Text2", "SpeechT5", "UniSpeech", "UniSpeech-SAT",
    "UnivNet", "VITS", "Wav2Vec2", "Wav2Vec2-BERT", "Wav2Vec2-Conformer",
    "Wav2Vec2Phoneme", "WavLM", "Whisper", "X-Codec", "XLS-R", "XLSR-Wav2Vec2"
  ],

  "video_models": [
    "SAM2 Video", "TimeSformer", "V-JEPA 2", "VideoMAE", "ViViT"
  ],

  "multimodal_models": [
    "ALIGN", "AltCLIP", "Aria", "AyaVision", "BLIP", "BLIP-2", "BridgeTower", "BROS",
    "Chameleon", "Chinese-CLIP", "CLIP", "CLIPSeg", "CLVP", "Cohere2Vision",
    "ColPali", "ColQwen2", "Data2Vec", "DeepseekVL", "DeepseekVLHybrid", "DePlot",
    "Donut", "EdgeTAM", "EdgeTamVideo", "Emu3", "Evolla", "FLAVA", "Florence2",
    "Gemma3", "Gemma3n", "GIT", "glm4v", "glm4v_moe", "GOT-OCR2", "GraniteVision",
    "Grounding", "DINO", "GroupViT", "IDEFICS", "Idefics2", "Idefics3",
    "InstructBLIP", "InstructBlipVideo", "InternVL", "Janus", "KOSMOS-2",
    "KOSMOS-2.5", "LayoutLM", "LayoutLMV2", "LayoutLMV3", "LayoutXLM",
    "LFM2-VL", "LiLT", "Llama4", "LLaVA", "LLaVA-NeXT", "LLaVa-NeXT-Video",
    "LLaVA-Onevision", "LXMERT", "MatCha", "MetaCLIP 2", "MGP-STR",
    "Mistral3", "mllama", "MM Grounding DINO", "Nougat", "OmDet-Turbo",
    "OneFormer", "Ovis2", "OWL-ViT", "OWLv2", "PaliGemma", "Perceiver",
    "PerceptionLM", "Phi4", "Multimodal", "Pix2Struct", "Pixtral",
    "Qwen2.5-Omni", "Qwen2.5-VL", "Qwen2Audio", "Qwen2VL",
    "Qwen3-Omni-MoE", "Qwen3VL", "Qwen3VLMoe", "ShieldGemma2",
    "SigLIP", "SigLIP2", "SmolLM3", "SmolVLM",
    "Speech Encoder Decoder Models", "TAPAS", "TrOCR", "TVLT", "TVP",
    "UDOP", "VideoLlava", "ViLT", "VipLlava",
    "Vision Encoder Decoder Models", "Vision Text Dual Encoder",
    "VisualBERT", "Voxtral", "X-CLIP"
  ],

  "reinforcement_learning_models": [
    "Decision Transformer", "Trajectory Transformer"
  ],

  "time_series_models": [
    "Autoformer", "Informer", "PatchTSMixer", "PatchTST",
    "Time Series Transformer", "TimesFM"
  ],

  "graph_models": [
    "Graphormer"
  ],

  "edge_cases": [
    "SignalP", "DeepMedic", "SecretomeP", "DeepLoc", "DeepLocPro"
  ]
}
\end{verbatim}

However, not all papers use these models. In the event that the paper does not
use one of these PTMs, identify the PTM.

The scientific method is made up of several steps. These include:

\begin{itemize}
    \item \textbf{Observation}: Identification of a phenomenon in the natural world.
    \item \textbf{Background}: Reviewing previous works and identifying relevant context.
    \item \textbf{Hypothesis}: Creating a testable explanation.
    \item \textbf{Test}: Experimenting and challenging the hypothesis.
    \item \textbf{Analysis}: Reviewing generated data and iterating on the hypothesis.
    \item \textbf{Conclusion}: Reporting results and supporting/rejecting the hypothesis.
\end{itemize}

\textbf{Objective}
Your task is to output only an array of JSON objects containing key-value pairs, where:
\begin{itemize}
    \item \texttt{model}: A string of the pre-trained deep learning model's name.
    \item \texttt{step}: The most probable step in the scientific process (listed above) where the model is leveraged.
    \item \texttt{prose}: The most salient excerpt from the paper showing concrete evidence of usage.
\end{itemize}
No explanations or extra output are allowed.

\textbf{Style}
Responses must be strictly machine-readable JSON. No natural language,
commentary, or formatting beyond the JSON object is permitted.

\textbf{Tone}
Neutral, objective, and machine-like.

\textbf{Audience}
The audience is a machine system that parses JSON. Human readability is
irrelevant.

\textbf{Response}
Return only an array of JSON objects of the form:

\begin{verbatim}
[
    {
        "model": string,
        "step": string,
        "prose": string
    },
    ...
]
\end{verbatim}

Nothing else should ever be returned.
\end{tcolorbox}
\clearpage    

\end{document}